\documentclass[journal]{IEEEtran}
\usepackage{amssymb}
\usepackage{graphicx}
\usepackage{caption}
\usepackage{booktabs}
\usepackage{amsmath}

\usepackage{multicol}
\usepackage{floatrow}
\usepackage{multirow}
\usepackage[table,xcdraw]{xcolor}
\usepackage{cite}
\usepackage[ruled,vlined]{algorithm2e}
\usepackage{algpseudocode}
\usepackage{epsfig}
\usepackage{epstopdf}
\usepackage[caption=false]{subfig}
\usepackage[normalem]{ulem}

\usepackage{amssymb}
\usepackage{pifont}

\usepackage[para,online,flushleft]{threeparttable}

\begin{document}
%


\title{Aerial-Aided mmWave VANETs Using NOMA: Performance Analysis, Comparison, and Insights}
%

\author{Abdullah~Abu~Zaid,
        Baha~Eddine~Youcef~Belmekki,
        and~Mohamed-Slim~Alouini
\thanks{Abdullah~Abu~Zaid, Baha~Eddine~Youcef~Belmekki, and Mohamed-Slim~Alouini are with the Computer, Electrical and Mathematical Science and Engineering Division, King Abdullah University of Science
and Technology (KAUST), Thuwal 23955-6900, Saudi Arabia (e-mail:
\{abdullah.abuzaid, bahaeddine.belmekki, slim.alouini\}@kaust.edu.sa). }}
%
%



\maketitle
\begin{abstract}
In this paper, we propose the integration of tethered flying platforms in cooperative vehicular ad hoc networks (VANETs) to alleviate the problems of rapid urbanization. In this context, we study the performance of VANETs by deriving approximate outage probability and average achievable rate expressions using tools from stochastic geometry. We compare between the usage of networked tethered flying platforms (NTFPs) and traditional roadside units (RSUs). On the other hand, the rapid increase of smart devices in vehicles and the upcoming urban air mobility (UAM) vision will congest the spectrum and require increased data rates. Hence, we use non-orthogonal multiple access (NOMA) to improve spectral efficiency and compare its performance to orthogonal access schemes. Furthermore, we utilize millimeter-wave (mmWave) frequencies for high data rates and implement a  sectored beamforming model. We extensively study the system using three transmission schemes: direct, relay, and hybrid transmission. The results show that when acting as relays, NTFPs outperform RSUs for larger distances between the transmitting and the receiving vehicles, while RSUs outperform NTFPs for short distances. However, NTFPs are the best solution when acting as a source. Moreover, we find that, in most cases, direct transmission is preferred to achieve a high rate compared to other schemes. Finally, the results are summarized in two tables that provide insights into connecting VANETs by selecting the most suitable platform and type of communication for a given set of parameters, configurations, and requirements.

\end{abstract}

\IEEEpeerreviewmaketitle

\section{Introduction}

\subsection{Motivation}
The deployment of intelligent devices in vehicles necessitates the development of reliable and secure vehicular communications. Vehicular ad hoc networks (VANETs) are dynamic networks that enable direct and cooperative communication between vehicles \cite{AdhocNetworksBook2001}, circumventing delay-inducing cellular networks \cite{hung2016efficient}. They are used to improve the safety of passengers and pedestrians, provide infotainment, and support intelligent systems like autonomous vehicles. Since vehicles are highly mobile and their antennas have a limited range, roadside units (RSUs) are used in VANETs to aid in relaying information. However, vehicles and RSUs suffer from intermittent connectivity in urban environments due to the plethora of objects that block line-of-sight (LOS) communications.

Non-terrestrial networks (NTNs) are flying platforms that leverage high altitudes to increase coverage and improve LOS conditions \cite{HAPS,huang2023system,extend}; they are considered a possible solution to enhance VANETs \cite{NTNs5G}. Free-flying NTNs, however, need to be recharged or refueled constantly and have limited wireless backhaul capacity \cite{giordani2020non}. To overcome these limitations, networked tethered flying platforms (NTFPs) utilize a tether connected to a ground unit, which supplies the flying platform with continuous power and a reliable backhaul link \cite{belmekkiNTFPs}. The advantages of NTFPs, such as a large coverage area, extended service times, reliability, and security, prompted researchers to study their use in communications \cite{kishk20203, alsamhi2020performance}.

Enhanced communication is not limited to ground vehicles alone. With the rapid growth of urban air mobility (UAM) anticipated in the next decade, it is considered the future of urban transportation \cite{UAM2018NASA}. UAM is recognized as the use of low-altitude unmanned aerial vehicles (UAVs) for package delivery and electric vertical take-off and landing (eVTOL) aircraft for passenger transportation \cite{zaid2023evtol}. Reliably connecting UAM aircraft is paramount to ensure the safety of passengers and pedestrians. Communication systems must be able to transmit critical information such as aircraft location, navigation data, and weather information \cite{UAM2018NASA}. Consequently, vehicles, NTFPs, and UAM aircraft must coexist and share limited resources with increased interference generated by UAM aircraft such as UAVs and eVTOLs. 

In recent years, research into smart cities has become widely popular. Inevitably, the number of communication devices in smart cities is expected to increase dramatically \cite{yaqoob2017enabling}. The existence of these devices will congest the spectrum in urban areas. Therefore, the use of non-orthogonal multiple access (NOMA) scheme has attracted much attention to improving spectral efficiency. NOMA is a multiple access scheme that allows users to transmit simultaneously on the same frequency and at the same time, but with different power levels \cite{saito2013system}. Moreover, some applications for future vehicles will require increased data rates; hence, a larger bandwidth. For instance, researchers and the industry are focusing on making fully autonomous vehicles a reality, which will require high data rates to operate safely \cite{ahangar2021survey}. Accordingly, higher frequency bands such as millimeter-wave (mmWave) and terahertz (THz) technologies are required to achieve high data rates. However, THz frequencies undergo high attenuation and are prone to blockage \cite{lou2023coverage}. Therefore, mmWave frequencies around $30$~GHz to $100$~GHz provides a large bandwidth that enables high data rates for vehicular communications.

\begin{table*}[]
\begin{threeparttable}
\caption{Selected related works.}
\label{tab:relatedworks}
\resizebox{\textwidth}{!}{%
\begin{tabular}{@{}cccccccc@{}}
\toprule
\textbf{Reference} & \textbf{Architecture} & \textbf{\begin{tabular}[c]{@{}c@{}}Communication \\ type\end{tabular}} & \textbf{\begin{tabular}[c]{@{}c@{}}Transmission\\ scheme\end{tabular}} & \textbf{Small-scale fading} & \textbf{Distribution models} & \textbf{\begin{tabular}[c]{@{}c@{}}MA scheme\\ Frequency\end{tabular}} & \textbf{Interference source} \\ \toprule
\cite{belmekki2021performance} & Ground & V2V/V2I & DT/RT & Rayleigh & \begin{tabular}[c]{@{}c@{}}Road intersection\\ Vehicles 1D PPP\end{tabular} & \begin{tabular}[c]{@{}c@{}}OMA\\ RF\end{tabular} & Other vehicles \\ \midrule
\cite{singh2022heterogeneous} & Ground & V2V & DT & \begin{tabular}[c]{@{}c@{}}Rayleigh \&\\ Lambertian\end{tabular} & \begin{tabular}[c]{@{}c@{}}Road intersection\\ Vehicles 1D PPP\end{tabular} & \begin{tabular}[c]{@{}c@{}}OMA\\ RF-VLC\end{tabular} & Other vehicles \\ \midrule
\cite{sun2020performance} & Ground & V2I & DT & Rayleigh & \begin{tabular}[c]{@{}c@{}}Roads PCP\\ Vehicles 1D PPP\end{tabular} & \begin{tabular}[c]{@{}c@{}}NOMA \& OMA\\ RF\end{tabular} & \begin{tabular}[c]{@{}c@{}}RSUs\\ Perfect SIC\end{tabular} \\ \midrule
\cite{belmekki2020performance} & Ground & V2V/V2I & RT & Rayleigh & \begin{tabular}[c]{@{}c@{}}Road intersection\\ Vehicles 1D PPP\end{tabular} & \begin{tabular}[c]{@{}c@{}}NOMA \& OMA\\ RF\end{tabular} & \begin{tabular}[c]{@{}c@{}}Other vehicles\\ Imperfect SIC\end{tabular} \\ \midrule
\cite{patel2021performance} & Ground & V2I & DT & \begin{tabular}[c]{@{}c@{}}Rayleigh \&\\ Nakagami-$m$\end{tabular} & No distribution model & \begin{tabular}[c]{@{}c@{}}NOMA\\ RF\end{tabular} & Imperfect SIC \\ \midrule
\cite{zhang2018analysis} & Ground & V2V/V2I & DT & Rayleigh & \begin{tabular}[c]{@{}c@{}}One-lane highway\\ Vehicles 1D PPP\end{tabular} & \begin{tabular}[c]{@{}c@{}}OMA\\ mmWave\\ No steering error\end{tabular} & \begin{tabular}[c]{@{}c@{}}Other vehicles\\ RSUs\end{tabular} \\ \midrule
\cite{ozpolat2018grid} & Ground & V2V & DT & Nakagami-$m$ & \begin{tabular}[c]{@{}c@{}}Manhattan roads\\ Vehicles 1D PPP\end{tabular} & \begin{tabular}[c]{@{}c@{}}OMA\\ mmWave\\ With steering error\end{tabular} & Other vehicles \\ \midrule
\cite{saluja2020energy} & Ground & V2I & DT & Rayleigh & \begin{tabular}[c]{@{}c@{}}Four-lane highway\\ Vehicles 1D PPP\end{tabular} & \begin{tabular}[c]{@{}c@{}}OMA\\ RF-mmWave\\ No steering error\end{tabular} & RSUs \\ \midrule
\cite{belmekki2020outage} & Ground & V2V/V2I & RT & \begin{tabular}[c]{@{}c@{}}Rayleigh \&\\ Nakagami-$m$\end{tabular} & \begin{tabular}[c]{@{}c@{}}Road intersection\\ Vehicles 1D PPP\end{tabular} & \begin{tabular}[c]{@{}c@{}}NOMA\\ mmWave\\ No steering error\end{tabular} & \begin{tabular}[c]{@{}c@{}}Other vehicles\\ No SIC\end{tabular} \\ \midrule
\cite{popoola4356704exploiting} & Aerial & V2I & DT & - & \begin{tabular}[c]{@{}c@{}}Rural roadway\\ SUMO simulator\end{tabular} & \begin{tabular}[c]{@{}c@{}}OMA\\ mmWave\\ No steering error\end{tabular} & Aerial platforms \\ \midrule
This work & Ground/Aerial & V2V/V2I & DT/RT/HT & \begin{tabular}[c]{@{}c@{}}Rayleigh \&\\ Nakagami-$m$\end{tabular} & \begin{tabular}[c]{@{}c@{}}Roads PCP\\ Vehicles 1D PPP\\ UAVs 2D BPP\end{tabular} & \begin{tabular}[c]{@{}c@{}}NOMA \& OMA\\ mmWave\\ With steering error\end{tabular} & \begin{tabular}[c]{@{}c@{}}Other vehicles*\\ Aerial platforms\\ Imperfect SIC\end{tabular} \\ \bottomrule
\end{tabular}
}
\begin{tablenotes}
\item[] \small * We note that our framework can easily incorporate RSUs as a source of interference into the analysis.
\end{tablenotes}
\end{threeparttable}
\end{table*}

\subsection{Related Works}
Recent performance-based research on VANETs that utilizes stochastic geometry focused on improving safety \cite{zhao2021interference} and maximizing throughput \cite{yang2021optimal}. To improve the connectivity of vehicles, researchers leveraged cooperative communications using RSUs \cite{belmekki2021performance, singh2022heterogeneous, wu2020modeling}. Authors in \cite{belmekki2021performance} showed that relay transmission outperforms direct transmission in intersections. In \cite{singh2022heterogeneous}, authors studied integrating radio frequency (RF) and visible light communications (VLC) in VANETs, where VLC is used for short distances, and RF is used for large distances. The study aims to serve future intelligent applications and improve safety. Several works in the literature used NOMA in VANETs \cite{belmekki2020performances,sun2020performance,belmekki2020performance, patel2021performance, bulut2022enhanced}. Authors in \cite{sun2020performance} derived analytical expressions for a NOMA vehicular network using a Poisson line cox process to model the vehicles and RSUs. In \cite{belmekki2020performance}, the authors showed that NOMA parameters such as the power coefficients and data rates should be chosen carefully to outperform OMA.  In \cite{patel2021performance}, authors studied the VANET performance of NOMA in a diversity scenario with non-identical channel distributions and showed results outperforming OMA. Authors in \cite{bulut2022enhanced} derived exact solutions of NOMA for a specific scenario in downlink vehicular networks. In addition, several works considered mmWave technology in VANETs \cite{zhang2018analysis, singh2021graphical, ozpolat2018grid, fatahi2020analytical, saluja2020energy, ahmed2023analyzing}. In \cite{zhang2018analysis}, the authors studied blockage and antenna effects on the performance of mmWave VANETs and showed that multi-hop networks outperform single-hop networks for high-distance communication. Authors in \cite{singh2021graphical} study a cluster-based VANET and show that the majority of the interference source is inter-cluster interference, and they proposed a scheme to minimize this source for enhanced connectivity. The authors in \cite{ozpolat2018grid} analyzed a mmWave VANET network in a grid-based urban environment to aid in designing and implementing such networks. They concluded that mmWave is capable of supporting communications in high-density scenarios. In \cite{fatahi2020analytical}, authors leveraged stochastic geometry to present a framework for mmWave vehicle-to-everything network, and provided results pertaining to caching performance. Authors in \cite{saluja2020energy} proposed a cooperative hybrid microwave-mmWave system for vehicular networks and obtained improved coverage results. In \cite{ahmed2023analyzing}, authors investigated the effect of Doppler shifts in mmWave VANETs, and showed degraded performance for high speeds. However, limited research has been done on the performance of NOMA in VANETs using mmWave technologies. The authors in \cite{belmekki2020outage} showed that the performance of NOMA in mmWave-aided vehicle communication at road junctions using non-line-of-sight (NLOS) transmission outperforms LOS transmission.
 
Research on integrating NTNs and NTFPs into communication networks was extensively investigated in the literature \cite{kishk20203, bushnaq2020optimal, popoola4356704exploiting, saluja2021design, xiao2023efficient, khabbaz2019modeling}. In \cite{kishk20203}, authors optimized the three-dimensional placement of NTFPs under the tether constraint to achieve minimum average path loss. While authors in \cite{bushnaq2020optimal} compared the performance of free-flying UAVs to tethered UAVs, and concluded that the latter would outperform the former given that the tether exceeds a given length. Authors in \cite{popoola4356704exploiting} studied a multi-array mmWave high altitude platform (HAP)-aided vehicular network, and concluded that there exists an optimal number of antenna elements to maximize coverage. In \cite{saluja2021design}, authors designed and analyzed an integrated aerial-terrestrial network to improve coverage and data rate performance. Authors in \cite{xiao2023efficient} proposed a novel architecture for data broadcasting in UAV-assisted VANETs, showing improved bandwidth efficiency. In \cite{khabbaz2019modeling}, authors studied a UAV-assisted vehicular network with a mobility model for the UAVs, and showed improved performance in terms of end-to-end connectivity.

To the best of our knowledge, no works investigated the benefits of NTFPs in VANETs; therefore, our paper aims to fill this gap. In addition, we conduct a comparative study between aerial-aided and terrestrial-aided networks in VANETs using NOMA and mmWave. We show the most suitable platform, multiple access scheme, and type of transmission used depending on the different scenarios and settings.

\subsection{Contribution}
The main contribution of this paper focuses on deriving performance metrics of an NTFP-aided VANET with the help of stochastic geometry. The paper derives the metrics for three transmission schemes and two relay configurations. Furthermore, the paper offers a comparative analysis between NTFPs and RSUs in VANETs and gives a framework for the best platform and transmission scheme choice, given the system parameters. Additionally, suitable models are implemented to best describe the mmWave channel. To summarize, key points of the contribution are given next:
\begin{itemize}
  \item Using tools from stochastic geometry, approximate outage probability (OP) and average achievable rate (AAR) expressions are derived, under three transmission schemes. Namely, direct transmission (DT), relay transmission (RT), and hybrid transmission (HT) schemes.
  \item Realistic distribution models for UAVs and ground vehicles are used. UAVs are modeled as a binomial point process (BPP), and ground vehicles are modeled as a Poisson cox process (PCP). Furthermore, the effect of increasing the density of interference sources is studied.
  \item The paper compares using an NTFP and an RSU for VANETs while considering the three transmission schemes. Furthermore, the system is studied when NTFPs/RSUs are acting as a relay and when acting as a source.
  \item Two LOS models are used in this paper that reflect the different types of links. An exponential LOS model is used for a ground-to-ground (G2G) link between two vehicles or an RSU, depending on the nodes' distance. On the other hand, a sigmoid-based LOS model is used for a ground-to-air (G2A) or air-to-ground (A2G) link, which is a function of the angle the link makes with the ground. Furthermore, a sectored antenna gain model is used to model the mmWave channel properly.
  \item This paper studies and compares the performance of NOMA and OMA in the context of VANETs. The behavior of NOMA is explored, and expressions for the outage rate and for the performance cross-point of NOMA versus OMA are derived, which signify the value at which NOMA stops outperforming OMA. The effect of imperfect successive interference cancellation (SIC) on the performance is also studied.
  \item Finally, the results of this paper are summarized by identifying the best choice of infrastructure and transmission scheme according to the data rate requirement, power allocation levels, priority ordering, distance of transmission, and configuration type. These insights can be used in VANETs to design mechanisms that opportunistically choose the most suitable platform, multiple access, and type of communication in concordance with system setup planning requirements.
\end{itemize}

\section{System Model}
\begin{figure}
    \centering
    \includegraphics[width=1\linewidth]{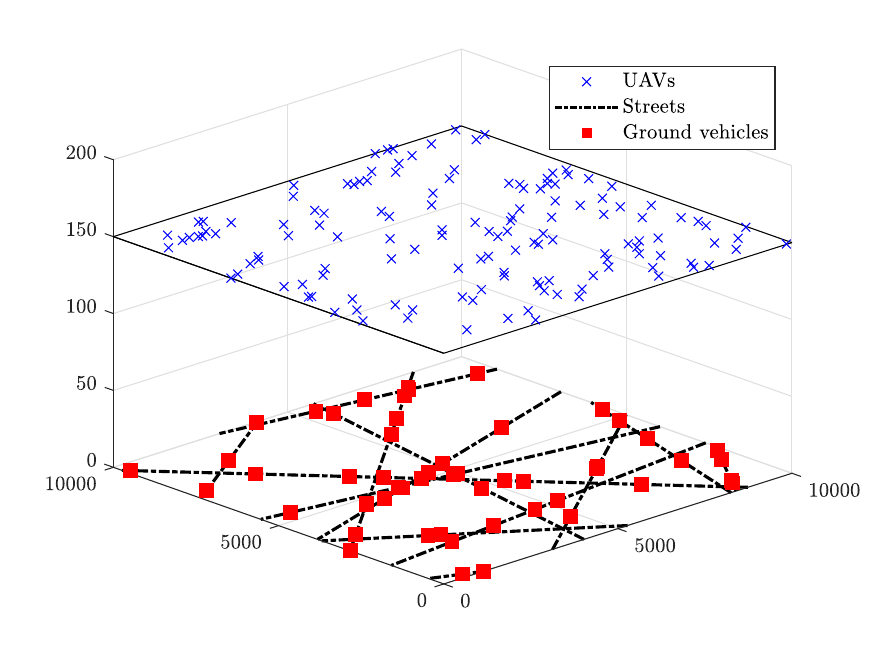}
    \caption{UAVs and ground vehicles distribution models.}
    \label{fig:SystemModelCoxBPP}
\end{figure}
\begin{figure*}
    \centering
    \includegraphics[width=0.7\linewidth]{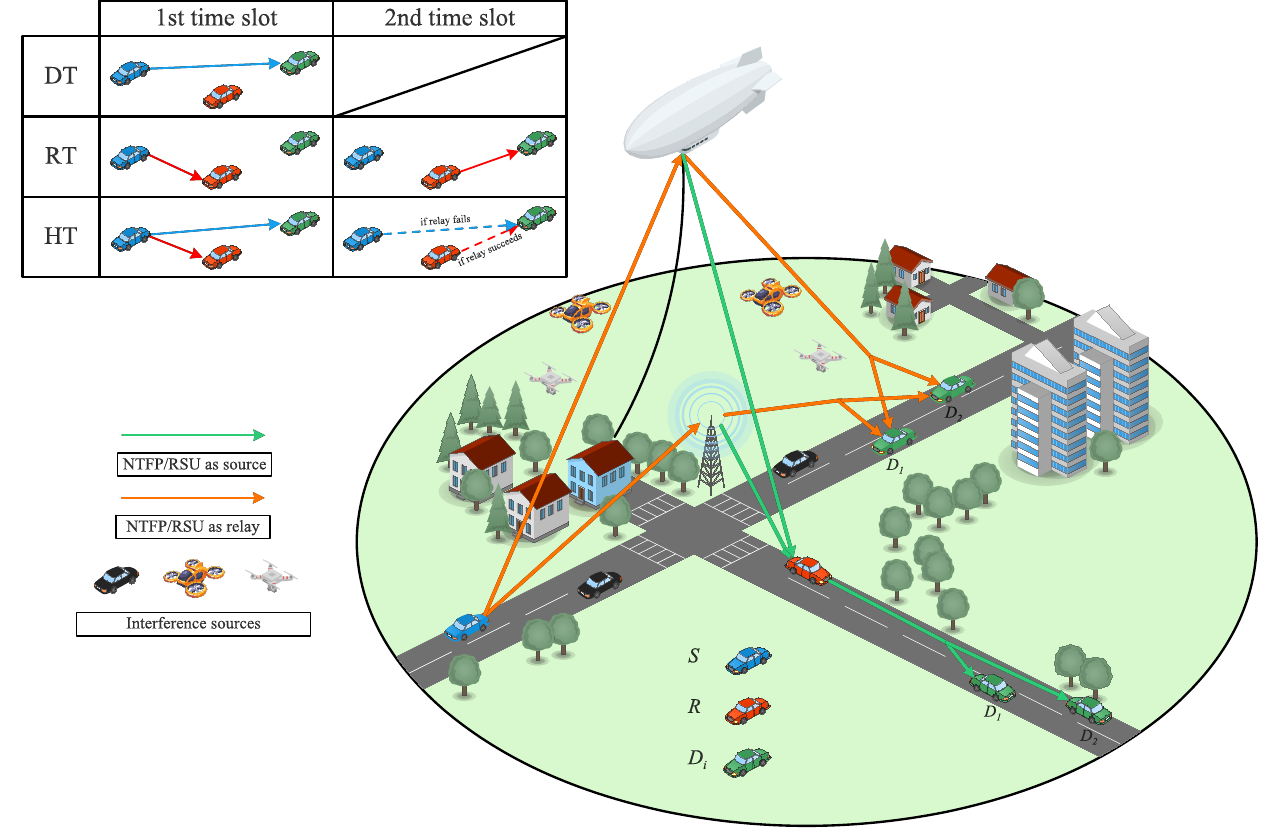}
    \caption{System model.}
    \label{fig:SystemModelFull}
\end{figure*}

\subsection{Environment}
We consider an urban environment with dense buildings. Two interference sources are assumed: ground vehicles (such as cars, trucks and motorcycles) and UAVs/eVTOLs. Note that for the remainder of the paper, eVTOLs are assumed to be UAVs. A fitting model used to model vehicles is the PCP, which is a doubly stochastic process composed of a Poisson line process (PLP) and a one-dimensional (1D) Poisson point process (PPP) \cite{2006PLPStreets}. The PLP models a street system as random lines, defined by an angle with respect to the horizontal axis and a perpendicular distance from the origin to the line. The 1D PPP models the positions of vehicles on each line. We denote the PLP by \(\Phi_L\) (with intensity \(\lambda_{\textrm{L}}\)), and denote the 1D-PPP by \(\Phi_{\textrm{V}}\) (with intensity \(\lambda_{\textrm{V}}\)). The receiving node is assumed to be on a fixed line denoted by \(L_0\). Hence, we denote the total interference from the PCP and \(L_0\) by \(\Phi_{L_t}=\Phi_L \cup \{L_0\}\). We model the UAVs, denoted by \(\Phi_{\textrm{U}}\), according to a two-dimensional (2D) binomial point process (BPP), which is more suitable than a PPP model for a UAV network \cite{chetlur2017BPP}. The UAVs are distributed on a disk with radius \(L\) and altitude~\(h_{\textrm{UAV}}\), centered around the origin. A simulated instance of the interference sources is shown in Fig.~\ref{fig:SystemModelCoxBPP}.

\subsection{Multiple Access Scheme}
We consider that the transmitting nodes use NOMA as a multiple access scheme to serve two users. There are two ordering configurations for the receiving nodes in NOMA: according to their channel status conditions \cite{ding2014performance} and quality of service (QoS) requirements \cite{QoSNOMA}. We consider the latter in this work since it is shown in \cite{QoSNOMA} that ordering the nodes with respect to their QoS requirements is more realistic. Consequently, we assume that user~$1$ requires low latency communication with a low data rate; hence, it has a high priority, and user~$2$ requires a high data rate and can accept a delay in transmission. For example, user~$1$ might be a vehicle receiving critical safety information that is small in size, and user~$2$ might be a passenger watching a video on stream.

\subsection{Transmission Schemes}
The density of urban environments often prohibits vehicles from communicating directly with other devices, which incentivizes the study of cooperative links using a relay. Hence, we study three transmission schemes: DT, RT, and HT. DT is when the source \(S\) sends directly to the destinations \(D_1\) and \(D_2\). It is the default scheme, which is done when the channel conditions are favorable. RT is when \(S\) sends to a relay \(R\) in the first time slot, then \(R\) uses decode and forward (DF) to forward the message to \(D_1\) and \(D_2\) in the second time slot. It has been shown that with NOMA, DF outperforms amplify and forward for low signal-to-noise ratio (SNR), and offers equivalent performance for high SNR \cite{jee2021coordinated}. The third scheme, HT, is a combination of DT and RT, where \(S\) sends to \(R\), \(D_1\), and \(D_2\) in the first time slot, then, in the second time slot, if \(R\) successfully decodes the message, it forwards it to \(D_1\) and \(D_2\), otherwise \(S\) sends the message again to \(D_1\) and \(D_2\). The transmission schemes are depicted in Fig.~\ref{fig:SystemModelFull}.

\subsection{Studied Configurations}
Two configurations are studied. First, NTFP/RSU as a relay, where a vehicle on the ground wants to communicate with another vehicle, possibly through an NTFP or an RSU \cite{belmekkiNTFPs}. An example is when vehicles want to send traffic and congestion information to other vehicles. Second, NTFP/RSU as a source, where an NTFP or RSU communicates a message to a vehicle. In this case, another vehicle may be used as a relay. This scenario may occur if an NTFP or an RSU wants to broadcast weather information to vehicles. The two configurations are shown in Fig.~\ref{fig:SystemModelFull}.




\subsection{Channel setting}
We assume that all nodes transmit with unit power while using mmWave frequencies. We consider an interference-limited channel, that is, the noise power is zero\footnote{We note that the impact of noise can be easily added into the analysis.}. The channel experiences a distance dependant path loss \(l_{pq}\) between nodes \(p \in \{S,R\}\) and \(q \in \{R,D_1,D_2\}\), \(p\neq q\), where \(l_{pq}=\|p-q\|^{-\alpha}\), and \(\alpha\) is the path loss exponent. We can denote the signal sent by \(S\) to \(D_1\) and \(D_2\) as
\begin{equation}
    \chi_S = \sqrt{a_1}\chi_{D_1} + \sqrt{a_2}\chi_{D_2},
\end{equation}
%
where \(a_1\) and \(a_2\) are the NOMA power coefficients for \(D_1\) and \(D_2\), respectively. We note that \(a_1+a_2=1\). The symbol \(\chi_{D_i}\) denotes the message intended for \(D_i\), where \(i\in\{1,2\}\).
\par
The signal received at \(q\) from \(p\) can be expressed as
\begin{equation}
    \begin{split}
        \mathcal{Y}_{pq} & =  h_{pq}\sqrt{l_{pq}}\chi_S+ \sum_{L\in\Phi_{L_t}}\sum_{v_{L}\in\Phi_{\textrm{V}_{q}}} h_{v_{L} q }\sqrt{l_{v_{L} q}}\chi_{v_{L}}\\
        & +\sum_{u\in\Phi_{{\textrm{U}}_{q}}} h_{uq}\sqrt{l_{uq}}\chi_u,
    \end{split}
\end{equation}

\noindent where \(\chi_{v_{L}}\) and \(\chi_u\) are the messages received from the vehicles and the UAVs, respectively. The coefficient \(h_{pq}\) represent small-scale fading between \(p\) and \(q\) that follows a Nakagami-$m$ random variable with parameter \(m\). We use Nakagami-$m$ since it is a general model that encompasses various channel conditions \cite{belmekki2019coooperative}.


The interference power from the vehicles and the UAVs to a node \(q\), denoted by \(I_{{V}_q}\) and \(I_{\textrm{U}_q}\), can now be defined as
\begin{equation}
    \begin{split}
        I_{{V}_q} & = I_{L_0}+I_{L} = \sum_{v_{L_0}\in\Phi_{\textrm{V}_{q}}}|h_{v_{L_0}q}|^2l_{v_{L_0}q}\\
        & + \sum_{L_j\in\Phi_{L}}\sum_{v_{L_j}\in\Phi_{\textrm{V}_{q}}} |h_{v_{L_j}q}|^2l_{v_{L_j}q},
    \end{split}
\end{equation}
where \(I_{L_0}\) and \(I_{L}\) are the interferences from the fixed line and the PCP, respectively, and
\begin{equation}
    I_{\textrm{U}_q} = \sum_{u\in\Phi_{{\textrm{U}}_{q}}} |h_{uq}|^2l_{uq}.
\end{equation}
%
\subsection{Probability of LOS}
Every link between two nodes in this system model can be either LOS or NLOS, where the path loss exponent and the Nakagami-$m$ parameter will be \(\alpha_{\textrm{LOS}}\) and \(m_{\textrm{LOS}}\) for a LOS link, and \(\alpha_{\textrm{NLOS}}\) and \(m_{\textrm{NLOS}}\) for a NLOS link, respectively. Next, we use two models to determine the probability of LOS for terrestrial and aerial links.

\subsubsection{Terrestrial Model}
A distance-based model is used for G2G links since it is shown to be a suitable model for urban blockages \cite{bai2014analysis}. The probability of LOS is given by
\begin{equation} \label{terrestPLOS}
    P_{\textrm{LOS}}^{(T)}\left(r\right) = e^{-\beta r},
\end{equation}
where \(r\) is the distance between the two nodes, and \(\beta\) is a parameter that characterises the density, shape, and heights of the buildings.

\subsubsection{Aerial Model}
An elevation-based model is used for A2G/G2A links. The authors in \cite{2003RECOMMENDATIONIP} proposed a sophisticated model for blockage probability in urban areas when a difference in altitudes is present. However, authors in \cite{al2014optimal} showed that a modified Sigmoid function closely approximates the model in \cite{2003RECOMMENDATIONIP} while offering mathematical tractability. The probability of LOS of this model is given by
\begin{equation}\label{aerialPLOS}
    P_{\textrm{LOS}}^{(A)}\left(r\right) = \frac{1}{1+a\textrm{exp}\left(-b\left(\theta(r)-a\right)\right)},
\end{equation}
where \(\theta(r)\) (in degrees) is the elevation angle between the two nodes, and \(a\) and \(b\) are parameters that reflect the density of buildings in the urban environment. The elevation angle is calculated as \(\theta(r) = \tan^{-1}\left(h/r\right)\), where \(h\) is the altitude of the aerial node, and \(r\) is the ground distance between the source and destination. Due to the rarity of blockage objects at high altitudes, the channel links between the UAVs and NTFP are assumed to be in LOS.

\subsection{Directional Beamforming}
Since mmWave frequencies require beamforming to be effective, we assume that all nodes have antenna arrays to enable directional beamforming. We use a sectored antenna model as in \cite{tractablebeamforming}. Therefore, the gain is maximum in the main lobe, denoted by \(G\), when the beam angle \(\phi\) is within a width \(\theta_{\textrm{BF}}\), i.e. \(|\phi| \leq \theta_{\textrm{BF}}\). On the other hand, the gain is minimum in the side lobes, denoted by \(g\), when \(|\phi| > \theta_{\textrm{BF}}\). Since it is difficult in practice to have perfect alignment of the transmitter and receiver antennas, we include beam steering errors for the main links \cite{wildman2014joint}. We assume the error, denoted by \(\epsilon\), is additive and follows a Gaussian distribution with mean zero and variance \({\sigma_{\textrm{e}}^{2}}\). Since the gain angle is symmetric, the absolute value of the beam steering error, \(|\epsilon|\), follows a half-normal distribution with cumulative distribution function (CDF) \(F_{|\epsilon|}(x)=\textrm{erf}(x/(\sqrt{2}\sigma_\textrm{e}))\).

The interfering links have beam angles that are distributed randomly according to a uniform random variable on \([0,2\pi]\). Hence, the gain of an interfering link is a discrete random variable given by
\begin{equation} \label{BeamformingGains}
    K = 
    \small
    \begin{cases}
    GG & \text{with prob. } P_{GG}=\left(\frac{\theta_{\textrm{BF}}}{2\pi}\right)^2,\\
    Gg & \text{with prob. } P_{Gg}= 2\left(\frac{\theta_{\textrm{BF}}}{2\pi}\right)\left(1-\frac{\theta_{\textrm{BF}}}{2\pi}\right),\\
    gg & \text{with prob. } P_{gg}=\left(1-\frac{\theta_{\textrm{BF}}}{2\pi}\right)^2.
    \end{cases}
    \normalsize
\end{equation}

\section{Outage Analysis}
In this section, we derive the approximate OP expressions with the aid of a tight bound given by \cite{alzer1997some}. First, we calculate the Laplace transforms of the interference sources required for the derivations. Second, we derive the OP using NOMA and for the different transmission schemes. Third, we derive the OP expressions when using OMA for comparison. Lastly, we find the total OP expressions considering the probability of LOS and beamforming models.

\subsection{Laplace Transform Expressions} \label{LaplaceExpressions}
In this part, we derive the Laplace transform expressions for the ground vehicles and the UAVs. We denote the Laplace transform for interference \(I_A\) by \(\mathcal{L}_{I_A}(s)\).

\subsubsection{Ground Vehicles}
We first calculate the Laplace transform of ground interference, \(I_V = I_{L_0}+I_L\). From the Laplace transform definition, we get \(\mathcal{L}_{I_V}(s)=\mathcal{L}_{I_{L_0}}(s)\mathcal{L}_{I_{L}}(s)\). Next, we calculate the Laplace transform expression of each source separately.
\newline
\indent \textbf{Lemma 1:} \textit{The Laplace transform of the interference from the vehicles on the line \(L_0\) is given by}
\begin{equation} \label{LT_IL0}
    \begin{split}
        \mathcal{L}_{I_{L_0}}(s) & = \textrm{exp}\!\left(\!-\lambda_{\textrm{V}}\int_{\mathbb{R}}\!1-\!\!\left(\frac{m}{s (x^2+z^2)^{-\frac{\alpha}{2}}+m}\right)^m\!\textrm{d}x\!\right),
    \end{split}
\end{equation}
\textit{where \(z\) is the height of the destination node.}
\newline
\textit{Proof:} See Appendix \ref{AppendixLemma1}.
\newline
\indent \textbf{Lemma 2:} \textit{The Laplace transform of the interference from the vehicles on all the other lines is given by}
\begin{equation} \label{LT_IL}
    \begin{split}
        \mathcal{L}_{I_{L}}(s) & = \textrm{exp}\Bigg[-2\pi\lambda_{\textrm{L}}\int_{\mathbb{R}^+}1-\textrm{exp}\Bigg(\!-\lambda_{\textrm{V}} \int_{\mathbb{R}}\!1\\
        & - \!\!\left(\frac{m}{s (x^2+y^2+z^2)^{-\frac{\alpha}{2}}+m}\right)^m\!\textrm{d}x\Bigg)\textrm{d}y\Bigg],
    \end{split}
\end{equation}
\textit{Proof:} See Appendix \ref{AppendixLemma2}.

\subsubsection{UAVs} \label{LT_IU}
Now we calculate the Laplace transform of the interference originating from the UAVs, \(I_U\), which follows a 2D BPP.
\newline
\indent \textbf{Lemma 3:} \textit{The Laplace transform of the interference from the UAVs is given by}
\begin{equation} \label{UAVLaplace}
    \begin{split}
        \mathcal{L}_{I_{\textrm{U}}}(s) = \left(\int_{\Delta h}^{\sqrt{{\Delta h}^2+L^2}}\left(\frac{m}{s u_x^{-\alpha}+m}\right)^m\frac{2u_x}{L^2}\textrm{d}u_x\right)^{N_{\textrm{U}}},
    \end{split}
\end{equation}
\textit{where \(\Delta h\) is the difference in heights between the receiver plane and the UAV plane, and} \(N_{\textrm{U}}\) \textit{is the number of UAVs.}
\newline
\textit{Proof:} See Appendix \ref{AppendixLemma3}.


\subsection{NOMA Outage Expressions}
Next, we proceed with deriving the expressions for OP when employing NOMA.

\subsubsection{Direct transmission} \label{NOMA_DT_OUTAGE}
First, We study the DT case, where \(S\) sends to \(D_1\) and \(D_2\) directly. We define the OP as the probability of the signal-to-interference ratio (SIR) falling below a threshold. The SIR at \(D_1\) to decode its intended message is given by
\begin{equation}
    \textrm{SIR}_{SD_{1\rightarrow 1}} = \frac{|h_{SD_1}|^2l_{SD_1}a_1}{|h_{SD_1}|^2l_{SD_1}a_2+I_{\textrm{tot}_{D_1}}},
\end{equation}
where \(I_{\textrm{tot}_{q}} = I_{{V}_{q}}+I_{\textrm{U}_{q}}\) denotes the total interference power received at node \(q\) from the vehicles and UAVs. In order for \(D_2\) to decode its intended message, it needs to decode the message of \(D_1\) and then remove it by applying SIC. Hence, the SIR for \(D_2\) to decode the message of \(D_1\) is given by
\begin{equation}
    \textrm{SIR}_{SD_{2\rightarrow 1}} = \frac{|h_{SD_2}|^2l_{SD_2}a_1}{|h_{SD_2}|^2l_{SD_2}a_2+I_{\textrm{tot}_{D_2}}},
\end{equation}
then, the SIR for \(D_2\) to decode its intended message is
\begin{equation}
    \textrm{SIR}_{SD_{2\rightarrow 2}} = \frac{|h_{SD_2}|^2l_{SD_2}a_2}{\gamma |h_{SD_2}|^2l_{SD_2}a_1 + I_{\textrm{tot}_{D_2}}},
\end{equation}
where \(\gamma\) is the remaining fraction of interference due to imperfect SIC.
Outage for \(D_1\) occurs if the SIR falls under the  threshold, \(T_1 = 2^{R_1}-1\), where \(R_1\) is the required rate for \(D_1\). Hence, we define the outage event for \(D_1\) as
\begin{equation}
    O_{D_1}^{(\mathbf{DT})} \triangleq [\textrm{SIR}_{SD_{1\rightarrow 1}}<T_1].
\end{equation}

Outage for \(D_2\) occurs if it fails to decode the message of \(D_1\) or it fails to decode its own message. Hence, the outage event is defined as 
\begin{equation}
    O_{D_2}^{(\mathbf{DT})} \triangleq [\textrm{SIR}_{SD_{2\rightarrow 1}}<T_1 \cup \textrm{SIR}_{SD_{2\rightarrow 2}}<T_2],
\end{equation}
where \(T_2 = 2^{R_2}-1\), and \(R_2\) is the required rate for \(D_2\). Next, we derive the outage probability, and we note that we employ an approximation using Alzer's inequality \cite{alzer1997some}, which is tight for small values of \(m\).

\textbf{Theorem 1:} \textit{The approximate OP for \(D_1\), denoted \(P(O_{D_1}^{(\mathbf{DT})})\), is given by}
\begin{equation}
    P(O_{D_1}^{(\mathbf{DT})}) \ { \cong}\  1-\sum_{k=0}^m (-1)^k \binom{m}{k} \mathcal{H}_{D_1}(ck\frac{mG_1}{l_{SD_1}}),
\end{equation}
\textit{where \(G_1=T_1/(a_1-T_1 a_2)\), and  \(\mathcal{H}_q(s)=\mathcal{L}_{I_{V}}(s)\mathcal{L}_{I_{\textrm{U}}}(s)\). The approximate OP for \(D_2\), denoted \(P(O_{D_2}^{(\mathbf{DT})})\), is given by}
\begin{equation}
    \begin{split}
        P(O_{D_2}^{(\mathbf{DT})}) & \ { \cong}\  1-\sum_{k=0}^m (-1)^k \binom{m}{k} \mathcal{H}_{D_2}(ck\frac{mG_{\textrm{max}}}{l_{SD_2}}),
    \end{split}
\end{equation}
\textit{where $G_{\textrm{max}}~=~\textrm{max}\{G_1, G_2\}$ \textit{and} $G_2~=~T_2/(a_2-\gamma T_2a_1)$.}
\newline
\textit{Proof:} See Appendix \ref{AppendixTheorem1}.

Note that \(D_1\) experiences an outage when \(T_1\geq\frac{a_1}{a_2}\). Hence, solving the inequality for the NOMA outage rate for DT, we get \({R^O_1}_{\textrm{DT}}=\textrm{log}_2(1+\frac{a_1}{a_2})\). Similarly, the NOMA outage rate for \(D_2\) using DT is \({R^O_2}_{\textrm{DT}}=\textrm{log}_2(1+\frac{a_2}{\gamma a_1})\).
\subsubsection{Relay transmission} \label{NOMA_RT_OUTAGE}
In the RT case, \(S\) sends to \(R\) in the first time slot, then \(R\) decodes the message of \(D_1\) and applies SIC to decode the message of \(D_2\). Next, \(R\) forwards both messages to \(D_1\) and \(D_2\), where \(D_1\) decodes its message, but \(D_2\) must first decode the message for \(D_1\) and then it can decode its intended message. In line with this, the outage events for \(D_1\) and \(D_2\) can be defined respectively as



\begin{equation}
    \begin{split}
        O_{{D_1}}^{(\mathbf{RT})} \triangleq [\textrm{SIR}_{SR_1}<T_1^{(\mathbf{RT})} 
        \cup \textrm{SIR}_{RD_{1\rightarrow 1}}<T_1^{(\mathbf{RT})}],
    \end{split}
\end{equation}
and
\begin{equation}
    \begin{split}
        O_{{D_2}}^{(\mathbf{RT})} & \triangleq [\textrm{SIR}_{SR_1}<T_1^{(\mathbf{RT})} \cup \textrm{SIR}_{SR_2}<T_2^{(\mathbf{RT})}\\
        &\ \cup \textrm{SIR}_{RD_{2\rightarrow1}}<T_1^{(\mathbf{RT})}
        \cup \textrm{SIR}_{RD_{2\rightarrow2}}<T_2^{(\mathbf{RT})}],
    \end{split}
\end{equation}
where \(T_i^{(\mathbf{RT})}=2^{2R_i}-1\), \(i\in\{1,2\}\). The rate \(R_i\) is multiplied by 2 since RT requires two time slots in comparison to DT.
\newline
\indent 
\textbf{Theorem 2:} \textit{The OP of \(D_1\) in the RT case is given by}
\begin{equation} \label{PoutD1_RT_NOMA}
    \begin{split}
        P(O_{D_1}^{(\mathbf{RT})}) &\ { \cong}\  1-\sum_{k=0}^m (-1)^k \binom{m}{k} \mathcal{H}_{R}\left(\frac{ckmG_1^{(\mathbf{RT})}}{l_{SR}}\right)\\
        & \times\sum_{k=0}^m (-1)^k \binom{m}{k} \mathcal{H}_{D_1}\left(\frac{ckmG_1^{(\mathbf{RT})}}{l_{RD_1}}\right),
    \end{split}
\end{equation}
\textit{where \(G_1^{(\mathbf{RT})}=T_1^{(\mathbf{RT})}/(a_1-T_1^{(\mathbf{RT})} a_2)\). The OP of \(D_2\) is given by}
\begin{equation} \label{PoutD2_RT_NOMA}
    \begin{split}
        P(O_{D_2}^{(\mathbf{RT})}) &\ { \cong}\  1-\sum_{k=0}^m (-1)^k \binom{m}{k}\mathcal{H}_{R_1}\left(\frac{ckmG_{\textrm{max}}^{(\mathbf{RT})}}{l_{SR}}\right)\\
        & \times\sum_{k=0}^m (-1)^k \binom{m}{k}\mathcal{H}_{D_2}\!\!\left(\frac{ckmG_{\textrm{max}}^{(\mathbf{RT})}}{l_{RD_2}}\right),
    \end{split}
\end{equation}
\textit{where we define the values} $G_{\textrm{max}}^{(\mathbf{RT})}~=~\textrm{max}\{G_1^{(\mathbf{RT})}, G_2^{(\mathbf{RT})}\}$ and $G_2~=~T_2^{(\mathbf{RT})}/(a_2-\gamma T_2^{(\mathbf{RT})}a_1)$.

\textit{Proof:} See Appendix \ref{AppendixTheorem2}.
\newline
The NOMA outage rate of \(D_1\) for RT happens when \(T_1^{(\mathbf{RT})}\geq\frac{a_1}{a_2}\). Solving for the NOMA outage rate, we get \({R^O_1}_{\textrm{RT}}~=~\frac{1}{2}\textrm{log}_2(1+\frac{a_1}{a_2})\). Similarly for \(D_2\), we obtain \({R^O_2}_{\textrm{RT}}~=~\frac{1}{2}\textrm{log}_2(1+\frac{a_2}{\gamma a_1})\).
\subsubsection{Hybrid transmission}
In the HT case, the outage event for \(D_1\) is a modified version of the event given in \cite{laneman2004cooperative}
\begin{equation}
    \begin{split}
        O_{D_1}^{(\mathbf{HT})} & \triangleq \Big[\textrm{SIR}_{SD_{1\rightarrow 1}}<\frac{T_1^{(\mathbf{RT})}}{2} \cap  \textrm{SIR}_{SR_1}<T_1^{(\mathbf{RT})}\Big]\\
        & \cup \ \Big[\textrm{SIR}_{SR_1}>T_1^{(\mathbf{RT})} \\
        & \cap  \textrm{SIR}_{RD_{1\rightarrow 1}}<T_1^{(\mathbf{RT})} \cap\textrm{SIR}_{SD_{1\rightarrow 1}}<T_1^{(\mathbf{RT})}\Big],
    \end{split}
\end{equation}
where \(\textrm{SIR}_{RD_{1\rightarrow 1}}\) and \(\textrm{SIR}_{SD_{1\rightarrow 1}}\) were not combined at the receiver due to intractability reasons. Simulations showed that this simplification does not affect the results. For \(D_2\), the outage event is defined in the same way as \(D_1\), which is given by
\begin{equation}
    \begin{split}
        O_{D_2}^{(\mathbf{HT})} & \triangleq \Big[(O_{SD_{2\rightarrow1}}^{(2)}\cup 
        O_{SD_{2\rightarrow2}}^{(2)}) \cap (O_{SR_1}\cup O_{SR_2})\Big]\\
        & \cup \Big[(O_{SR_1}^C\cap O_{SR_2}^C) \cap (O_{RD_{2\rightarrow1}}\cup O_{RD_{2\rightarrow2}})\\
        & \cap (O_{SD_{2\rightarrow1}}\cup O_{SD_{2\rightarrow2}})\Big],
    \end{split}
\end{equation}

\noindent where \(O_{SD_{2\rightarrow i}}^{(2)}=[\textrm{SIR}_{SD_{2\rightarrow i}}<~\frac{1}{2}T_i^{(\mathbf{RT})}]\).


\begin{figure*}[!t]
\normalsize
\begin{equation} \label{Pout_NOMA_HT_D1}
\begin{split}
        P(O_{D_1}^{(\mathbf{HT})}) &\ { \cong}\  1-\sum_{k=0}^m (-1)^k \binom{m}{k} \mathcal{H}_{D_1}\left(\frac{ckm\frac{T_1^{(\mathbf{RT})}}{2}}{l_{SD_1}(a_1-a_2\frac{T_1^{(\mathbf{RT})}}{2})}\right)- \sum_{k=0}^m (-1)^k \binom{m}{k} \mathcal{H}_{R}\left(\frac{ckmG_1^{(\mathbf{RT})}}{l_{SR}}\right)\\
        & + \Bigg[\sum_{k=0}^m (-1)^k \binom{m}{k} \mathcal{H}_{D_1}\left(\frac{ckm\frac{T_1^{(\mathbf{RT})}}{2}}{l_{SD_1}(a_1-a_2\frac{T_1^{(\mathbf{RT})}}{2})}\right) \times \sum_{k=0}^m (-1)^k \binom{m}{k} \mathcal{H}_{R}\left(\frac{ckmG_1^{(\mathbf{RT})}}{l_{SR}}\right)\Bigg]\\
        & + \sum_{k=0}^m (-1)^k \binom{m}{k} \mathcal{H}_{R}\left(\frac{ckmG_1^{(\mathbf{RT})}}{l_{SR}}\right) \times\Bigg[1-\sum_{k=0}^m (-1)^k \binom{m}{k} \mathcal{H}_{D_1}\left(\frac{ckmG_1^{(\mathbf{RT})}}{l_{RD_1}}\right)\Bigg]\\ 
        & \times\Bigg[1-\sum_{k=0}^m (-1)^k \binom{m}{k} \mathcal{H}_{D_1}\left(\frac{ckmG_1^{(\mathbf{RT})}}{l_{SD_1}}\right)\Bigg].
    \end{split}
\end{equation}

\begin{equation} \label{Pout_NOMA_HT_D2}
    \begin{split}
        \!\!\!\!\!\!\!\!\!\!\!\!\!\!\!\! P(O_{D_2}^{(\mathbf{HT})}) &\ { \cong}\  1-\sum_{k=0}^m (-1)^k \binom{m}{k} \mathcal{H}_{D_2}\left(\frac{ckm\mathcal{Z}}{l_{SD_2}}\right)-\sum_{k=0}^m (-1)^k \binom{m}{k} \mathcal{H}_{R}\left(\frac{ckmG_{\textrm{max}}^{(\mathbf{RT})}}{l_{SR}}\right)\\
        & +\Bigg[\sum_{k=0}^m (-1)^k \binom{m}{k} \mathcal{H}_{D_2}\left(\frac{ckm\mathcal{Z}}{l_{SD_2}}\right)\times \sum_{k=0}^m (-1)^k \binom{m}{k} \mathcal{H}_{R}\left(\frac{ckmG_{\textrm{max}}^{(\mathbf{RT})}}{l_{SR}}\right)\Bigg]\\
        & +\sum_{k=0}^m (-1)^k \binom{m}{k} \mathcal{H}_{R}\left(\frac{ckmG_{\textrm{max}}^{(\mathbf{RT})}}{l_{SR}}\right)\times \Bigg[1-\sum_{k=0}^m (-1)^k \binom{m}{k} \mathcal{H}_{D_2}\left(\frac{ckmG_{\textrm{max}}^{(\mathbf{RT})}}{l_{RD_2}}\right)\Bigg]\\
        & \times \Bigg[1-\sum_{k=0}^m (-1)^k \binom{m}{k} \mathcal{H}_{D_2}\left(\frac{ckmG_{\textrm{max}}^{(\mathbf{RT})}}{l_{SD_2}}\right)\Bigg].
    \end{split}
\end{equation}
\hrulefill
\vspace*{4pt}
\end{figure*}

\indent 
\textbf{Theorem 3:} \textit{The OP of \(D_1\) in the HT case is given by Eq. \eqref{Pout_NOMA_HT_D1}, and the OP of \(D_2\) is given by Eq. \eqref{Pout_NOMA_HT_D2} in the top of the next page, where}

\begin{equation}
    \mathcal{Z}=\textrm{max}\Bigg\{\frac{\frac{1}{2}T_1^{(\mathbf{RT})}}{a_1-\frac{1}{2}T_1^{(\mathbf{RT})}a_2}, \frac{\frac{1}{2}T_2^{(\mathbf{RT})}}{a_2-\gamma\frac{1}{2}T_2^{(\mathbf{RT})}a_1}\Bigg\}.
\end{equation}

\textit{Proof:} See Appendix \ref{AppendixTheorem3}.
\newline
The NOMA outage rate of \(D_1\) for HT occurs when \(T_1^{(\mathbf{RT})}\geq\frac{2a_1}{a_2}\). Solving for the NOMA outage rate, we get \({R^O_1}_{\textrm{HT}}~=~\frac{1}{2}\textrm{log}_2(1+\frac{2a_1}{a_2})\). Similarly for \(D_2\), we have \({R^O_2}_{\textrm{HT}}~=~\frac{1}{2}\textrm{log}_2(1+\frac{2a_2}{\gamma a_1})\)
\subsection{OMA Outage Expressions}
The derivation of the OP when using OMA follows the same steps as in NOMA. However, since the access scheme is orthogonal, there is no cross-interference between \(D_1\) and \(D_2\). Due to space limitations, we only calculate the OP for DT. In the DT case, the SIR at \(D_i\) to decode its message is

\begin{equation}
    \textrm{SIR}_{SD_{i\rightarrow i}}^{(\textrm{OMA})} = \frac{|h_{SDi}|^2l_{SD_i}}{I_{\textrm{tot}_{D_i}}}.
\end{equation}

To calculate the OP of \(D_i\), we follow the same steps as in Appendix~\ref{AppendixLemma1}, hence, we obtain
\begin{equation}
    \begin{split}
        P(O_{D_i}^{(\mathbf{DT}-\textrm{OMA})}) & \ { \cong}\ 1-\sum_{k=0}^m (-1)^k \binom{m}{k} \mathcal{H}_{D_i}(ck\frac{m\Theta_{i}}{l_{SD_i}}),
    \end{split}
\end{equation}
where \(\Theta_i=2^{2R_i}-1\). The rate \(R_i\) is multiplied by 2 since OMA uses double the resources of NOMA. We note that for RT and HT we have \(\Theta_i^{(\mathbf{RT})}=2^{4R_i}-1\).
\subsection{Total Outage Expression} \label{total_outage}
We derive and present the total OP expressions considering the probability of LOS and the beamforming models. We start with the LOS model (LSM).

\subsubsection{Outage with LOS model}
The LSMs are considered on the main links and interfering links. The probability density function (PDF) of the fading links, considering LSM, is defined as follows
\begin{equation}
    f_h(x) \triangleq P_{\textrm{LOS}}(r_x)f_{h_{\textrm{LOS}}}(x)+(1-P_{\textrm{LOS}}(r_x))f_{h_{\textrm{NLOS}}}(x),
\end{equation}
where \(f_{h_{\textrm{LOS}}}(x)\) is the LOS fading PDF with parameter \(m_{\textrm{LOS}}\), and \(f_{h_{\textrm{NLOS}}}(x)\) is the NLOS fading PDF with parameter \(m_{\textrm{NLOS}}\). Note that  \(P_{\textrm{LOS}}(r_x)\) can be the terrestrial model defined in Eq.~\eqref{terrestPLOS}, or the aerial model defined in Eq.~\eqref{aerialPLOS}. Hence, the OP with LSM is given by
\begin{equation}
    P_{O_{\textrm{LSM}}} = P_{\textrm{LOS}}(r_x)P_{O_{\textrm{LOS}}}+(1-P_{\textrm{LOS}}(r_x))P_{O_{\textrm{NLOS}}},
\end{equation}
where \(P_{O_{\textrm{LOS}}}\) is the OP with parameters \(m_{\textrm{LOS}}\) and \(\alpha_{\textrm{LOS}}\), and \(P_{O_{\textrm{NLOS}}}\) is the OP with parameters \(m_{\textrm{NLOS}}\) and \(\alpha_{\textrm{NLOS}}\). Next, we calculate the Laplace transforms of the interferences given this fading model. We derive the expressions for a general \(P_{\textrm{LOS}}(r)\). The Laplace transform of the interference source \(I_{L_0}\) considering LSM is found as follows
\begin{equation} \label{LT_IL0_LSM}
    \begin{split}
        \mathcal{L}_{I_{L_0}}^{\textrm{(LSM)}}(s) & = \mathbb{E}\left[\prod_{v\in{L_0}}\textrm{exp}\left(-s|h_{v q}|^2l_{v q}\right)\right] \\
        &\!\!\!\!\!\!\!\!\!\!\!\!\!\!\!\!\!\!\!\!\!\!\!\!\! =\mathbb{E}\Bigg[\prod_{v\in{L_0}}\!\!\!\Big(P_{\textrm{LOS}}(r)\mathbb{E}_{h_{v q\sim \textrm{LOS}}}\left[\textrm{exp}(-s|h_{v q}|^2l_{v q})\right]\\
        &\!\!\!\!\!\!\!\!\!\!\!\!\!\!\!\!\!\!\!\!\!\!\!\!\! + (1-P_{\textrm{LOS}}(r))\mathbb{E}_{h_{v q\sim \textrm{NLOS}}}\left[\textrm{exp}(-s|h_{v q}|^2l_{v q})\right]\Big)\Bigg]\\
        &\!\!\!\!\!\!\!\!\!\!\!\!\!\!\!\!\!\!\!\!\!\!\!\!\! \stackrel{(a)}{=} \textrm{exp}\Bigg(-\lambda_{\textrm{V}}\!\!\int_{\mathbb{R}}1-\left(\frac{1}{\frac{s r^{-\alpha_{\textrm{NLOS}}}}{m_{\textrm{NLOS}}}+1}\right)^{m_{\textrm{NLOS}}}\!\!\!\!\!\!- P_{\textrm{LOS}}(r)\\
        &\!\!\!\!\!\!\!\!\!\!\!\!\!\!\!\!\!\!\!\!\!\!\!\!\! \times \Bigg[\left(\frac{1}{\frac{s r^{-\alpha_{\textrm{LOS}}}}{m_{\textrm{LOS}}}+1}\right)^{m_{\textrm{LOS}}}\!\!\!\!\!\! - \left(\frac{1}{\frac{s r^{-\alpha_{\textrm{NLOS}}}}{m_{\textrm{NLOS}}}+1}\right)^{m_{\textrm{NLOS}}}\Bigg]\textrm{d}r\Bigg),
    \end{split}
\end{equation}
where in \((a)\), the inner expectations are computed as in Appendix~\ref{AppendixLemma1}, then we use the PGFL of a 1D PPP. The Laplace transform of the interference source \(I_{L}\) considering LSM is calculated by following the same steps in Appendix~\ref{AppendixLemma2}; hence, we obtain
\begin{equation}
    \begin{split}
        \mathcal{L}_{I_{L}}^{\textrm{(LSM)}}(s) & = \textrm{exp}\Bigg[-2\pi\lambda_{\textrm{L}}\int_{\mathbb{R}^+}1-\textrm{exp}\Bigg(-\lambda_{\textrm{V}} \int_{\mathbb{R}}1\\
        &\!\!\!\!\!\!\!\!\!\!\!\!\!\!\!\!\!\!\! -\left(\frac{m_{\textrm{NLOS}}}{s (x^2+y^2+z^2)^{-\frac{\alpha_{\textrm{NLOS}}}{2}}+m_{\textrm{NLOS}}}\right)^{m_{\textrm{NLOS}}}\\
        &\!\!\!\!\!\!\!\!\!\!\!\!\!\!\!\!\!\!\! + P_{\textrm{LOS}}(r)\Bigg[\left(\frac{m_{\textrm{LOS}}}{s (x^2+y^2+z^2)^{-\frac{\alpha_{\textrm{LOS}}}{2}}+m_{\textrm{LOS}}}\right)^{m_{\textrm{LOS}}}\\
        &\!\!\!\!\!\!\!\!\!\!\!\!\!\!\!\!\!\!\! - \left(\frac{m_{\textrm{NLOS}}}{s (x^2+y^2+z^2)^{-\frac{\alpha_{\textrm{NLOS}}}{2}}+m_{\textrm{NLOS}}}\right)^{m_{\textrm{NLOS}}}\Bigg]\textrm{d}x\Bigg)\textrm{d}y\Bigg].
    \end{split}
\end{equation}
Similarly, the Laplace transform of \(I_{U}\) considering LSM is obtained using the results of Appendix~\ref{AppendixLemma3}, to get
\begin{equation}
    \begin{split}
        \mathcal{L}_{I_U}^{\textrm{(LSM)}}(s) & = \Bigg(\int_{\Delta h}^{\sqrt{{\Delta h}^2+L^2}}\left(\frac{m_{\textrm{NLOS}}}{s u_x^{-\alpha_{\textrm{NLOS}}}+m_{\textrm{NLOS}}}\right)^{m_{\textrm{NLOS}}}\\
        & + P_{\textrm{LOS}}(u_x)\Bigg[\left(\frac{m_{\textrm{LOS}}}{s u_x^{-\alpha_{\textrm{LOS}}}+m_{\textrm{LOS}}}\right)^{m_{\textrm{LOS}}}\\
        & -\left(\frac{m_{\textrm{NLOS}}}{s u_x^{-\alpha_{\textrm{NLOS}}}+m_{\textrm{NLOS}}}\right)^{m_{\textrm{NLOS}}}\Bigg]\frac{2u_x}{L^2}\textrm{d}u_x\Bigg)^{N_{\textrm{U}}}.\\
    \end{split}
\end{equation}

\subsubsection{Outage with beamforming models}
We present the total OP with the LOS model and beamforming model. First, we show the effect of the beam steering error on the main links. It can be shown that the total OP in the presence of beam steering errors is given by \cite{turgut2017coverage}
\begin{equation}
    \begin{split}
        P_{O_{\textrm{tot}}}&=F_{|\epsilon|}(\theta_{\textrm{BF}}/2)^2P_{O_{\textrm{LSM}}}(GG)\\
        & +2F_{|\epsilon|}(\theta_{\textrm{BF}}/2)(1-F_{|\epsilon|}(\theta_{\textrm{BF}}/2))P_{O_{\textrm{LSM}}}(Gg)\\
        & +(1-F_{|\epsilon|}(\theta_{\textrm{BF}}/2))^2P_{O_{\textrm{LSM}}}(gg),
    \end{split}
\end{equation}
where \(P_{O_{\textrm{LSM}}}(K)\) is the OP considering LSM with gain \(K\), and \(\theta_{\textrm{BF}}\) is the beam width of the main lobe. Finally, we show the Laplace transforms considering the beamforming model. Since \(K\) is a discrete random variable with three values as in Eq.~\eqref{BeamformingGains}, the interference source \(I_A\) will be composed of three parts, and by using the thinning property, the three processes will be independent. Hence, the interference can be written as \cite{turgut2017coverage}
\begin{equation}
    I_A^{\textrm{(tot)}}=I_{A, GG}^{\textrm{(LSM)}}+I_{A, Gg}^{\textrm{(LSM)}}+I_{A, gg}^{\textrm{(LSM)}}=\sum_{K}I_{A, K}^{\textrm{(LSM)}},
\end{equation}
where \(K=\{GG,Gg,gg\}\). The thinning theorem states that each process will have a density equal to \(\lambda_A P_K\), where \(P_K\) is given in Eq.~\eqref{BeamformingGains}. Hence, the Laplace transform for \(I_A^{\textrm{(tot)}}\) is given by
\begin{equation}
    \begin{split}
        \mathcal{L}_{I_A^{\textrm{(tot)}}}(s)&=\mathbb{E}_{I_A^{\textrm{(tot)}}}\left[\textrm{e}^{-s{I_A}^{\textrm{(tot)}}}\right]\\ 
        & =\mathbb{E}_{I_A^{\textrm{(tot)}}}\left[e^{-s(I_{A, GG}^{\textrm{(LSM)}}+I_{A, Gg}^{\textrm{(LSM)}}+I_{A, gg}^{\textrm{(LSM)}})}\right]\\
        & =\prod_K \mathbb{E}_{I_{A, K}^{\textrm{(LSM)}}}\left[e^{-sI_{A, K}^{\textrm{(LSM)}}}\right]\\ 
        & =\prod_K \mathcal{L}_{I_{A, K}^{\textrm{(LSM)}}}(s).
    \end{split}
\end{equation}

\subsection{NOMA Data Rate and Power Coefficient} \label{NOMA_a1_derivation}
In this section, we derive the cross-point values at which the performance of NOMA and OMA are equal in DT and RT. Hence, to ensure that NOMA has better performance, the power coefficient, \(a_1\), and the data rates must be chosen carefully. Furthermore, we provide the range of the SIC parameter \(\gamma\) such that NOMA outperform OMA for \(D_2\).

\subsubsection{User-1}
To find the values for which NOMA outperform OMA for \(D_1\) in the DT case, we write
\begin{equation}
    P(O_{D_1}^{(\mathbf{DT})}) < P(O_{D_1}^{(\mathbf{DT}-\textrm{OMA})}).
\end{equation}

Which is equivalent to
\begin{equation} \label{DT_NOMA_OMA_Outage_equation}
    G_1 < \Theta_1, \ \textrm{or}\ \frac{2^{R_1}-1}{a_1-a_2(2^{R_1}-1)} < 2^{2R_1}-1.
\end{equation}

Solving for the rate, we get
\begin{equation}
    R_1 < \textrm{log}_2\left(\frac{1+\sqrt{1-4a_1(1-a_1)}}{2(1-a_1)}\right),
\end{equation}
and we define the point at which the performance of NOMA and OMA cross in DT as \({R^c_1}_{\textrm{DT}}\). Hence, given \(a_1\), the rate \(R_1\) has to be chosen less than \({R^c_1}_{\textrm{DT}}\) in order for NOMA to outperform OMA. Solving for \(a_1\) in Eq. \eqref{DT_NOMA_OMA_Outage_equation}, we get 
\begin{equation}
    a_1 > \frac{T_1(1+\Theta_1)}{\Theta_1(1+T_1)},
\end{equation}
and, given \(R_1\), we define the cross-point as \({a^c_1}_{\textrm{DT}}\). The same steps are followed in the RT case to arrive at
\begin{equation}
    R_1 < \frac{1}{2}\textrm{log}_2\left(\frac{1+\sqrt{1-4a_1(1-a_1)}}{2(1-a_1)}\right),
\end{equation}
while solving for \(a_1\) gives
\begin{equation}
    a_1 > \frac{T_1^{(\mathbf{RT})}(1+\Theta_{1}^{(\mathbf{RT})})}{\Theta_{1}^{(\mathbf{RT})}(1+T_1^{(\mathbf{RT})})}.
\end{equation}

We define the cross-points for RT as \({R^c_1}_{\textrm{RT}}\) and \({a^c_1}_{\textrm{RT}}\). Since HT is a combination of DT and RT, its cross-point values will be between the cross-point values of DT and RT.
\subsubsection{User-2}
For \(D_2\), the following inequality should be satisfied in the DT case to achieve better performance than OMA
\begin{equation}
        \textrm{max}\big\{G_1, G_2\big\} < \Theta_2.
\end{equation}

Solving for \(a_1\), we get
\begin{equation} \label{NOMA_vs_OMA_a1}
    \begin{cases}
         a_1 < \frac{\Theta_2-T_2}{\Theta_2(1+\gamma T_2)}, & \textrm{if } G_1<G_2,\vspace{0.2cm}\\
         
         a_1 > \frac{T_1(1+\Theta_2)}{\Theta_2(1+T_1)}, & \textrm{if } G_1>G_2.\\
    \end{cases}
\end{equation}

We can solve for \(R_2\) in both cases to obtain
\begin{equation}
    R_2 >
    \begin{cases}
         \textrm{log}_2\left(\frac{1-a_1-\sqrt{(a_1-1)^2-4a_1^2\gamma(1-\gamma)}}{2a_1\gamma}\right), & \textrm{if } G_1<G_2,\\
         \frac{1}{2} \textrm{log}_2\left(\frac{a_2\ 2^{R_1}}{(a_1-1)2^{R_1}+1}\right), & \textrm{if } G_1>G_2.\\
    \end{cases}
\end{equation}

We define the cross-points as \({R^c_2}_{\textrm{DT}}\) and \({a^c_2}_{\textrm{DT}}\).  Finally, using Eq.~\eqref{NOMA_vs_OMA_a1}, the range of values of \(\gamma\) that enable NOMA to outperform OMA is given by

\begin{equation}
    \gamma < \frac{\Theta_2-T_2-a_1\Theta_2}{a_1\Theta_2T_2},
\end{equation}
where we define \(\gamma^c_{\textrm{DT}}\) to be the cross-point. The expressions for RT can be obtained following the same steps, therefore, they are omitted.

\section{Average Achievable Rate Analysis}
In this section, we derive the AAR when using NOMA, and for the different transmission schemes.


\subsection{Direct Transmission} 
We first start by calculating the AAR using DT. We define the AAR for \(D_i\) as
\begin{equation}
    \mathcal{R}_{D_i}^{\mathbf{(DT)}} \triangleq \mathbb{E}[\textrm{log}_2(1+\textrm{SIR}_{SD_{i\rightarrow i}}^{\textrm{(tot)}})],
\end{equation}
where \(\textrm{SIR}_{pq}^{\textrm{(tot)}}\) is the total SIR, taking into account LSM and the beamforming models. To compute the AAR, we make use of the following relation for the expectation of a positive random variable~\(X\)
\begin{equation}
    \mathbb{E}[X] = \int_{u>0} P(X>u)\textrm{d}u.
\end{equation}
Hence, we get
\begin{equation}
    \begin{split}
        \mathcal{R}_{D_i}^{\mathbf{(DT)}} & = \int_{u>0} \mathbb{E}_{I}\Big[\mathbb{P}(\textrm{log}_2(1+\textrm{SIR}_{SD_{i\rightarrow i}}^{\textrm{(tot)}}) > u)\Big]\textrm{d}u\\
        & = \int_{u=0}^{t_i} \mathbb{E}_{I}\Big[\mathbb{P}(\textrm{SIR}_{SD_{i\rightarrow i}}^{\textrm{(tot)}}>2^u-1)\Big]\textrm{d}u,
    \end{split}
\end{equation}
where \(t_1 = \textrm{log}_2(1+\frac{a_1}{a_2})\) and \(t_2 = \textrm{log}_2(1+\frac{a_2}{\gamma a_1})\). Note that for perfect SIC, we have \(t_2=\infty\). The expectation inside the integral is equivalent to the success probability, given by \(P_{S_{\textrm{tot}, SD_i}}=1-P_{O_{\textrm{tot}, SD_i}}\). To compute \(P_{S_{\textrm{tot}, SD_i}}\) we follow the same steps as in Appendix~\ref{AppendixLemma3}, then we use the results from section \ref{total_outage} to get the total expression as

\begin{equation}
    \begin{split}
    \mathcal{R}_{D_i}^{\mathbf{(DT)}} & = \int_{u=0}^{t_i} P_{S_{\textrm{tot}, SD_i}}\textrm{d}u = \int_{u=0}^{t_i}F_{|\epsilon|}(\theta_{\textrm{BF}}/2)^2 P_{S_{\textrm{LSM}, SD_i}}(GG)\\
    & +2F_{|\epsilon|}(\theta_{\textrm{BF}}/2)(1-F_{|\epsilon|}(\theta_{\textrm{BF}}/2)) P_{S_{\textrm{LSM}, SD_i}}(Gg)\\
    & +(1-F_{|\epsilon|}(\theta_{\textrm{BF}}/2))^2 P_{S_{\textrm{LSM}, SD_i}}(gg) \textrm{d}u,
    \end{split}
\end{equation}
where
\begin{equation} \label{PS_LSM_SD_eq}
    \begin{split}
        P_{S_{\textrm{LSM}, SD_i}}(g_1 g_2) & \cong P_{\textrm{LOS}}({l_{SD_i}}^{\alpha_{\textrm{LOS}}})\Bigg(\sum_{k=0}^{m_{\textrm{LOS}}} (-1)^k \binom{m_{\textrm{LOS}}}{k}\\
        &\!\!\!\!\!\!\!\!\!\!\!\!\!\!\!\!\!\!\!\!\!\!\!\!\!\!\!\! \times \mathcal{H}_{D_i}^{\textrm{(tot)}}\left(\frac{g_1 ckm_{\textrm{LOS}}(2^u-1)}{g_2 l_{SD_i} (a_1-a_2(2^u-1))}\right)\Bigg)\\
        &\!\!\!\!\!\!\!\!\!\!\!\!\!\!\!\!\!\!\!\!\!\!\!\!\!\!\!\!  + (1-P_{\textrm{LOS}}({l_{SD_i}}^{\alpha_{\textrm{LOS}}})) \Bigg(\sum_{k=0}^{m_{\textrm{NLOS}}} (-1)^k \binom{m_{\textrm{NLOS}}}{k}\\
        &\!\!\!\!\!\!\!\!\!\!\!\!\!\!\!\!\!\!\!\!\!\!\!\!\!\!\!\! \times \mathcal{H}_{D_i}^{\textrm{(tot)}}\left(\frac{g_1 ckm_{\textrm{NLOS}}(2^u-1)}{g_2 l_{SD_i} (a_1-a_2(2^u-1))}\right)\Bigg),
    \end{split}
\end{equation}
and
\begin{equation}
        \mathcal{H}_{D_i}^{\textrm{(tot)}}(s) = \prod_K \mathcal{L}_{I_{L_0, K}^{\textrm{(LSM)}}}(s) \cdot \prod_K \mathcal{L}_{I_{L, K}^{\textrm{(LSM)}}}(s) \cdot \prod_K \mathcal{L}_{I_{U, K}^{\textrm{(LSM)}}}(s).
\end{equation}

\subsection{Relay Transmission}
When using RT, the AAR will be the minimum AAR of both transmission links. Hence, we define the AAR for \(D_i\) as
\begin{equation}
    \begin{split}
        \mathcal{R}_{D_i}^{\mathbf{(RT)}}\!\! \triangleq \mathbb{E}[\textrm{min}\{&\frac{1}{2}\textrm{log}_2(1+\textrm{SIR}_{SR_i}^{(\textrm{tot})}),\\
        & \frac{1}{2}\textrm{log}_2(1+\textrm{SIR}_{RD_{i\rightarrow i}}^{(\textrm{tot})})\}].
    \end{split}
\end{equation}

To calculate the AAR for \(D_i\), we proceed as follows
\begin{equation}
    \begin{split}
        \mathcal{R}_{D_i}^{\mathbf{(RT)}} & = \mathbb{E}[\frac{1}{2}\textrm{log}_2(1+\textrm{min}\{\textrm{SIR}_{SR_i}^{(\textrm{tot})},\textrm{SIR}_{RD_{i\rightarrow i}}^{(\textrm{tot})}\})]\\
        &\!\!\!\!\!\!\!\!\!\!\!\!\!\!\!\!\!\! = \int_{u=0}^{f_i} \mathbb{E}_{I}\Big[\mathbb{P}(\textrm{min}\{\textrm{SIR}_{SR_i}^{(\textrm{tot})}, \textrm{SIR}_{RD_{i\rightarrow i}}^{(\textrm{tot})}\} > T_u)\Big]\textrm{d}u\\
        &\!\!\!\!\!\!\!\!\!\!\!\!\!\!\!\!\!\! = \int_{u=0}^{f_i} \mathbb{E}_{I}\Big[ \mathbb{P}(\textrm{SIR}_{SR_i}^{(\textrm{tot})}>T_u)\Big] \times\mathbb{E}_{I}\Big[\mathbb{P}(\textrm{SIR}_{RD_{i\rightarrow i}}^{(\textrm{tot})}>T_u)\Big]\textrm{d}u\\
        &\!\!\!\!\!\!\!\!\!\!\!\!\!\!\!\!\!\!  = \int_{u=0}^{f_i} P_{S_{\textrm{tot}, SR_i}} P_{S_{\textrm{tot}, RD_i}}\textrm{d}u,
    \end{split}
\end{equation}
where \(T_u=2^{2u}-1\). The LSM success probabilities are given by
\begin{equation} \label{PS_LSM_SR_eq}
    \begin{split}
        P_{S_{\textrm{LSM}, SR_i}}(g_1 g_2) & \cong P_{\textrm{LOS}}({l_{SR_i}}^{\alpha_{\textrm{LOS}}})\\
        & \!\!\!\!\!\!\!\!\!\!\!\!\!\!\!\!\!\!\!\!\!\!\!\!\!\!\!\!\!\!\!\!\!\!\!\!\!\!\!\!\!\!\! \times \Bigg(\sum_{k=0}^{m_{\textrm{LOS}}} (-1)^k \binom{m_{\textrm{LOS}}}{k}\mathcal{H}_{R_i}^{\textrm{(tot)}}\left(\frac{g_1 ckm_{\textrm{LOS}}  T_u}{g_2 l_{SR_i} (a_1-a_2  T_u)}\right)\Bigg)\\
        & \!\!\!\!\!\!\!\!\!\!\!\!\!\!\!\!\!\!\!\!\!\!\!\!\!\!\!\!\!\!\!\!\!\!\!\!\! + (1-P_{\textrm{LOS}}({l_{SR_i}}^{\alpha_{\textrm{LOS}}}))\\
        & \!\!\!\!\!\!\!\!\!\!\!\!\!\!\!\!\!\!\!\!\!\!\!\!\!\!\!\!\!\!\!\!\!\!\!\!\!\!\!\!\!\!\! \times \Bigg(\sum_{k=0}^{m_{\textrm{NLOS}}} (-1)^k \binom{m_{\textrm{NLOS}}}{k}\mathcal{H}_{R_i}^{\textrm{(tot)}}\left(\frac{g_1 ckm_{\textrm{NLOS}}  T_u}{g_2 l_{SR_i} (a_1-a_2  T_u)}\right)\Bigg),
    \end{split}
\end{equation}
and
\begin{equation} \label{PS_LSM_RD_eq}
    \begin{split}
        P_{S_{\textrm{LSM}, RD_i}}(g_1 g_2) & \cong P_{\textrm{LOS}}({l_{RD_i}}^{\alpha_{\textrm{LOS}}})\\
        & \!\!\!\!\!\!\!\!\!\!\!\!\!\!\!\!\!\!\!\!\!\!\!\!\!\!\!\!\!\!\!\!\!\!\!\!\!\!\!\!\!\!\! \times \Bigg(\sum_{k=0}^{m_{\textrm{LOS}}} (-1)^k \binom{m_{\textrm{LOS}}}{k}\mathcal{H}_{D_i}^{\textrm{(tot)}}\left(\frac{g_1 ckm_{\textrm{LOS}}  T_u}{g_2 l_{RD_i} (a_1-a_2  T_u)}\right)\Bigg)\\
        & \!\!\!\!\!\!\!\!\!\!\!\!\!\!\!\!\!\!\!\!\!\!\!\!\!\!\!\!\!\!\!\!\!\!\!\!\! + (1-P_{\textrm{LOS}}({l_{RD_i}}^{\alpha_{\textrm{LOS}}}))\\
        & \!\!\!\!\!\!\!\!\!\!\!\!\!\!\!\!\!\!\!\!\!\!\!\!\!\!\!\!\!\!\!\!\!\!\!\!\!\!\!\!\!\!\! \times \Bigg(\sum_{k=0}^{m_{\textrm{NLOS}}} (-1)^k \binom{m_{\textrm{NLOS}}}{k}\mathcal{H}_{D_i}^{\textrm{(tot)}}\left(\frac{g_1 ckm_{\textrm{NLOS}}  T_u}{g_2 l_{RD_i} (a_1-a_2  T_u)}\right)\Bigg).
    \end{split}
\end{equation}
    
where \(f_1 = \frac{1}{2}\textrm{log}_2(1+\frac{a_1}{a_2})\), and \(f_2 = \frac{1}{2}\textrm{log}_2(1+\frac{a_2}{\gamma a_1})\). Note that when \(\gamma=0\), we have \(f_2=\infty\).

\subsection{Hybrid Transmission}
The AAR when using HT depends on \(R\) if it successfully decodes the messages. Hence, it is given by the following equation for \(D_i\)
\begin{equation}
    \mathcal{R}_{D_i}^{\mathbf{(HT)}} \triangleq
    \small
    \begin{cases}
    \underbrace{
    \begin{split}
         \mathbb{E}\Big[\textrm{max}\Big\{\textrm{min}\{&\frac{1}{2}\textrm{log}_2(1+\textrm{SIR}_{SR_i}^{(\textrm{tot})}), \\
         & \frac{1}{2}\textrm{log}_2(1+\textrm{SIR}_{RD_{i\rightarrow i}}^{(\textrm{tot})})\}, \\
         & \frac{1}{2}\textrm{log}_2(1+\textrm{SIR}_{SD_{i\rightarrow i}}^{(\textrm{tot})})\Big\}\Big],
    \end{split}}_{\text{if \(R\) decodes}}\\
    \underbrace{\mathbb{E}\Big[\frac{1}{2}\textrm{log}_2(1+2\textrm{SIR}_{SD_{i\rightarrow i}}^{(\textrm{tot})})\Big].}_{\text{if \(R\) doesn't decode}}\\
    \end{cases}
    \normalsize
\end{equation}
The AAR when \(R\) does not decode the message is obtained the same way as the AAR for DT. In the case when \(R\) decodes the message, the AAR is given by
\begin{equation}
    \begin{split}
        \mathcal{R}_{D_i}^{\mathbf{(HT)}} 
        & = \int_{u=0}^{f_i} \mathbb{E}_{I}\Big[ \mathbb{P}(\textrm{SIR}_{SR_i}^{(\textrm{tot})}>T_u)\Big]\mathbb{E}_{I}\Big[\mathbb{P}(\textrm{SIR}_{RD_{i\rightarrow i}}^{(\textrm{tot})}>T_u)\Big]\\
        &+ \mathbb{E}_{I}\Big[\mathbb{P}(\textrm{SIR}_{SD_{i\rightarrow i}}^{(\textrm{tot})}>T_u)\Big]- \Big(\mathbb{E}_{I}\Big[ \mathbb{P}(\textrm{SIR}_{SR_i}^{(\textrm{tot})}>T_u)\Big]\\
        &\times \mathbb{E}_{I}\Big[\mathbb{P}(\textrm{SIR}_{RD_{i\rightarrow i}}^{(\textrm{tot})}>T_u)\Big] \mathbb{E}_{I}\Big[\mathbb{P}(\textrm{SIR}_{SD_{i\rightarrow i}}^{(\textrm{tot})}>T_u)\Big]\Big)\textrm{d}u\\
        & = \int_{u=0}^{f_i} P_{S_{\textrm{tot}, SR_i}} P_{S_{\textrm{tot}, RD_i}} + P_{S_{\textrm{tot}, SD_i}}\\
        &\ \ \ \ \ \ \  - P_{S_{\textrm{tot}, SR_i}} P_{S_{\textrm{tot}, RD_i}} P_{S_{\textrm{tot}, SD_i}}\textrm{d}u,
    \end{split}
\end{equation}
where \(P_{S_{\textrm{tot}, SD_i}}\) is given by \eqref{PS_LSM_SD_eq} but the rate is multiplied by \(2\). \(P_{S_{\textrm{tot}, SR_i}}\) and \(P_{S_{\textrm{tot}, RD_i}}\) are given by \eqref{PS_LSM_SR_eq} and \eqref{PS_LSM_RD_eq}, respectively.

\begin{table}[h!]
\footnotesize
\centering\caption{Simulation parameters.} \label{simulation_parameters}
\begin{tabular}{|c|c|c|c|c|c|}
\hline
\textbf{Symbol}            & \textbf{Value} & \textbf{Symbol}          & \textbf{Value} & \textbf{Symbol}          & \textbf{Value}          \\ \hline
\hline
\(R_1\)                    & 0.5 bits/s     & \(m_{\textrm{NLOS}}\)    & 1              & \(h_{\textrm{UAV}}\)     & 150 m                   \\ \hline
\(R_2\)                    & 1.5 bits/s     & \(a\)                    & 11.95          & \(h_{\textrm{NTFP}}\)    & 500 m                   \\ \hline
\(a_1\)                    & 0.8            & \(b\)                    & 0.136          & \(G\)                    & 63                      \\ \hline
\(a_2\)                    & 0.2            & \(\beta\)                 & 0.0095         & \(g\)                    & 0.63                    \\ \hline
\(\alpha_{\textrm{LOS}}\)  & 2              & \(\lambda_{\textrm{L}}\) & 1e-3           & \(\theta_{\textrm{BF}}\) & \(30^{\circ}\)          \\ \hline
\(\alpha_{\textrm{NLOS}}\) & 4              & \(\lambda_{\textrm{V}}\) & 5e-4           & \(\sigma_e\)             & \(10^{\circ}\)          \\ \hline
\(m_{\textrm{LOS}}\)       & 2              & \(N_{\textrm{U}}\)       & 500 UAVs       & \(L\)                    & \(10^4\) m \\ \hline
\end{tabular}
\end{table}

\begin{figure}[ht!]
    \centering
    \includegraphics[width=1\linewidth]{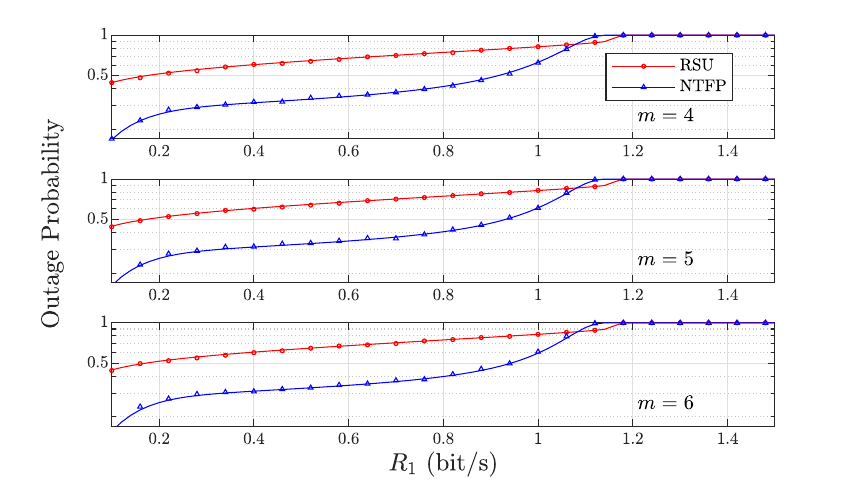}
    \captionof{figure}{The accuracy of the derivations shown with simulation for \(m=4,5,6\).}
    \label{fig:Poutvsm_4_5_6}
\end{figure}

\section{Simulation Results and Discussion}
In this section, we evaluate the performance of our system model. We show the performance in terms of OP and AAR. We assume the simulation area is \(20\) km by \(20\) km (\(400~\textrm{km}^2\)). Unless otherwise stated, the figures study the NTFP/RSU as a relay using RT, with the values of the simulation parameters given in Table~\ref{simulation_parameters}. Furthermore, we assume perfect SIC, i.e. \(\gamma=0\), unless otherwise stated. Without loss of generality, we set the distances between \(S\) and destinations as \(||S-D_1||=220\)~m and \(||S-D_2||=230\)~m and assume \(R\) is located in the midpoint. To validate our analysis, we perform Monte-Carlo simulations, where the marks in the figures represent the simulation results.

\begin{figure}
    \subfloat[\(D_1\).]{%
      \includegraphics[clip,width=0.48\linewidth]{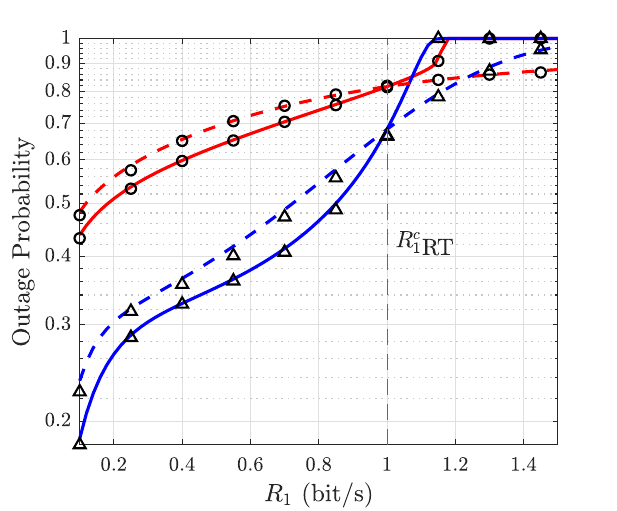}\label{fig:PoutvsR1_NOMA_OMA}
    }
    \subfloat[\(D_2\).]{%
      \includegraphics[clip,width=0.48\linewidth]{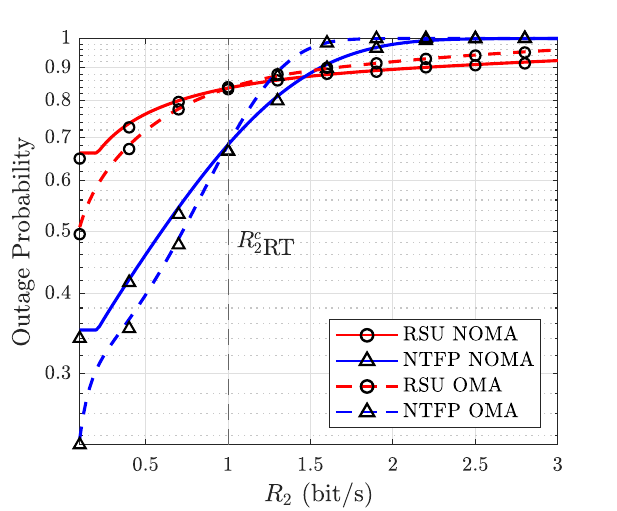}\label{fig:PoutvsR2_NOMA_OMA}
    }
    \caption{\(P_{\textrm{out}}\) as a function of data rates using NOMA and OMA.}
\end{figure}

\begin{figure}[t]
    \subfloat[\(P_{O_{\textrm{LOS}}}\) and \(P_{O_{\textrm{NLOS}}}\) for \(D_1\).\label{1a}]{%
    \includegraphics[width=0.45\linewidth]{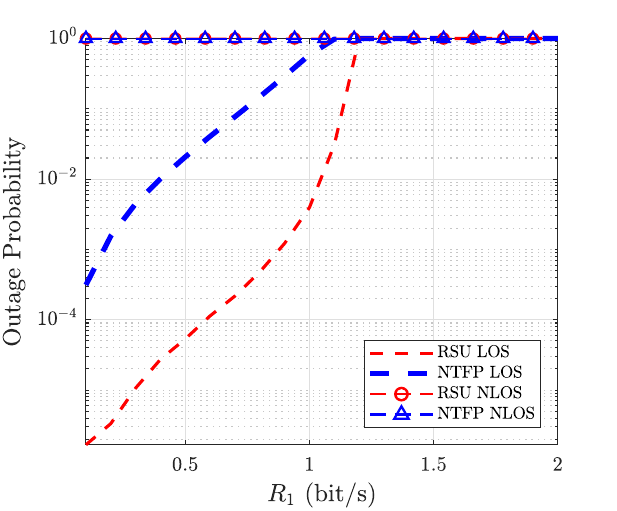}}
    \subfloat[\(P_{O_{\textrm{LSM}}}\) for \(D_1\).\label{1b}]{%
    \includegraphics[width=0.45\linewidth]{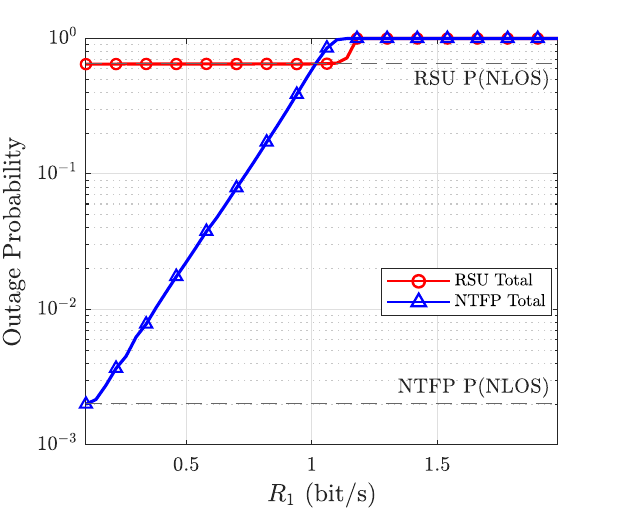}}
    \\
    \subfloat[\(P_{O_{\textrm{LOS}}}\) and \(P_{O_{\textrm{NLOS}}}\) for \(D_2\).\label{1c}]{%
    \includegraphics[width=0.45\linewidth]{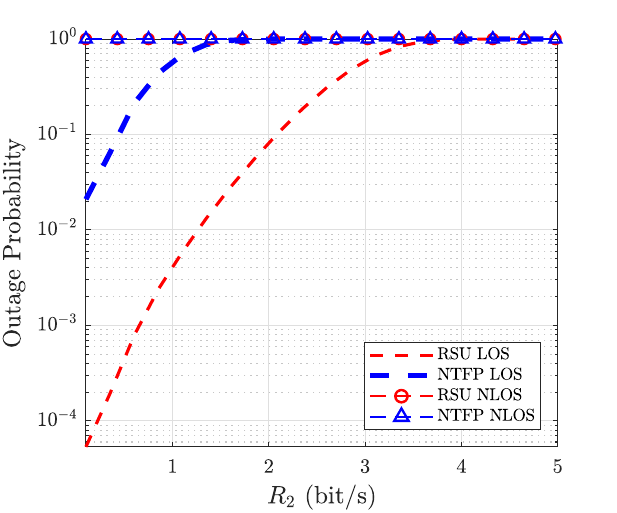}}
    \subfloat[\(P_{O_{\textrm{LSM}}}\) for \(D_2\).\label{1d}]{%
    \includegraphics[width=0.45\linewidth]{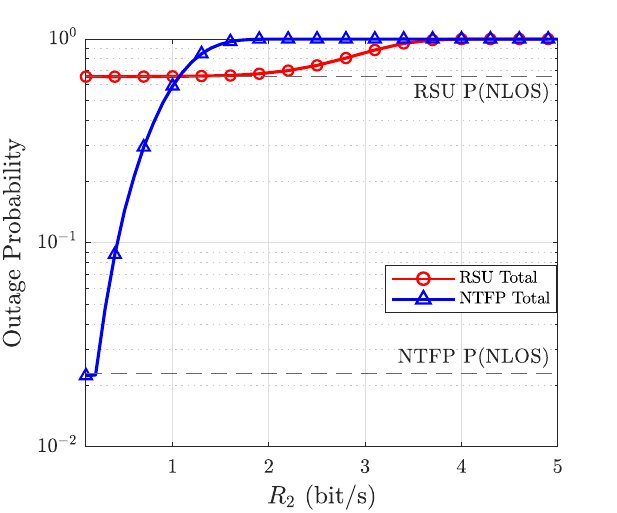}}
    \caption{Impact of the LOS probability on performance.}
    \label{fig:LOS_NLOS_Total} 
\end{figure}

In Fig.~\ref{fig:PoutvsR1_NOMA_OMA} and Fig.~\ref{fig:PoutvsR2_NOMA_OMA}, we plot the OP of \(D_1\) and \(D_2\) as a function of their respective data rates. We can see from Fig.~\ref{fig:PoutvsR1_NOMA_OMA} that NOMA outperforms OMA for  \(R_1<1\)~bit/s. However, for \(R_1>1\)~bit/s, NOMA experiences an outage for \(D_1\), which makes OMA superior for higher rates. This is because, in this setting, NOMA is set to provide \(D_1\) with low latency and low data rate. In Fig.~\ref{fig:PoutvsR2_NOMA_OMA}, we notice that OMA outperforms NOMA for small values of \(R_2\). This is because \(D_2\) must decode the message of \(D_1\) before decoding its message. Furthermore, the performance of NOMA is constant at the start; the reason for this is that decoding of \(D_2\) message depends only on \(R_1\) for small values of \(R_2\). Increasing \(R_2\) improves the performance of NOMA over OMA. This is due to OMA using twice the resources of NOMA. At high data rate requirements, we see that the RSU outperforms the NTFP; this is explained in Fig.~\ref{fig:LOS_NLOS_Total}, where we show why the RSU in some cases outperforms the NTFP.

Fig.~\ref{fig:LOS_NLOS_Total} draws \(P_{O_{\textrm{LOS}}}\), \(P_{O_{\textrm{NLOS}}}\), and the combined link \(P_{O_{\textrm{LSM}}}\). Due to the larger distance between \(S\) and the NTFP compared to the RSU, the LOS link of the RSU will experience less attenuation, as seen in Fig.~\ref{1a} and Fig.~\ref{1c}. The NLOS links of both RSU and NTFP will also experience strong attenuation at large distances, which explains the high OP. Furthermore, since the NTFP has a high altitude, it serves the majority of nodes with a LOS link. These facts combined will result in the performance shown in Fig.~\ref{1b} and Fig.~\ref{1d}.


\begin{figure*}[ht!]
    \centering
    \subfloat[\label{fig:PoutvsSD1Distance_NTFPHeight}]{%
    \includegraphics[width=0.25\linewidth]{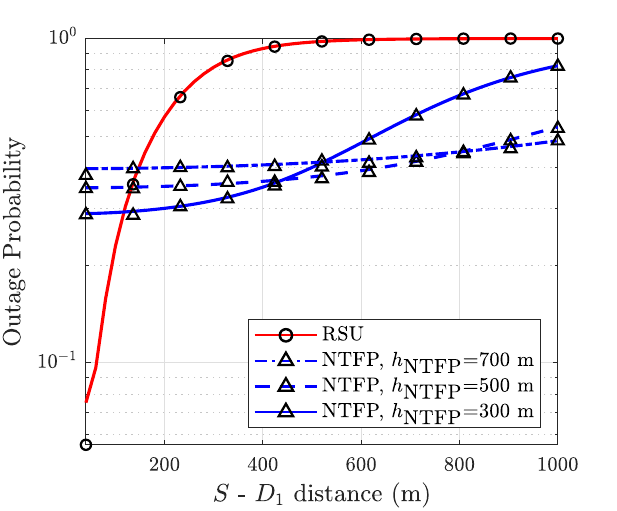}}
    \subfloat[\label{fig:AARvsSD1Distance_NTFPHeight}]{%
    \includegraphics[width=0.25\linewidth]{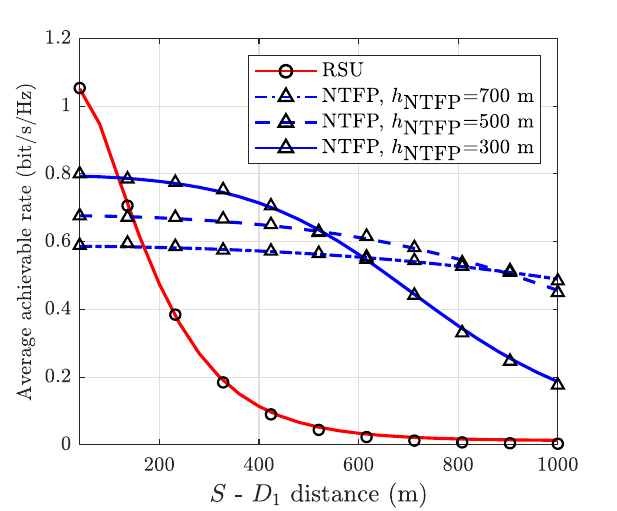}}
    \subfloat[\label{fig:PoutvsSDdist_DT_RT_HT}]{%
    \includegraphics[width=0.25\linewidth]{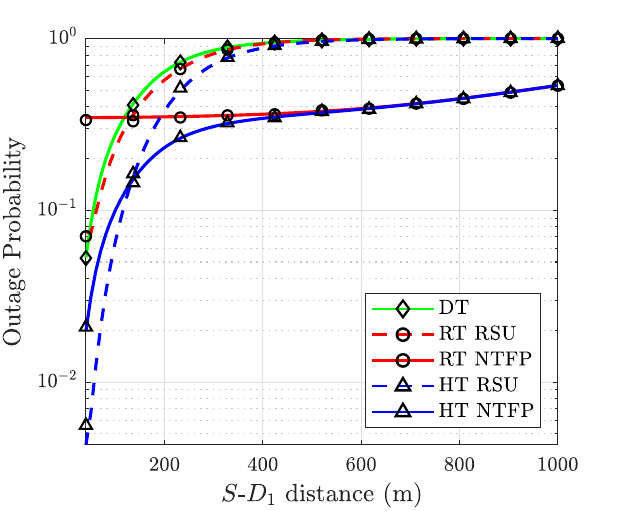}}
    \subfloat[\label{fig:AARvsSDdist_DT_RT_HT}]{%
    \includegraphics[width=0.25\linewidth]{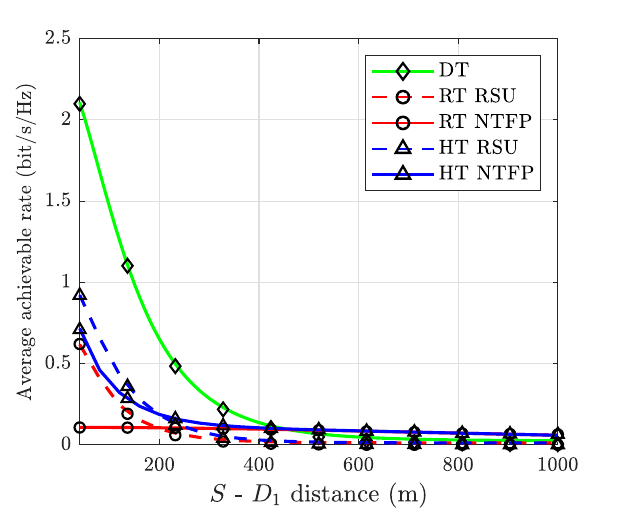}}
    \caption{Performance as a function of source-destination distance. (a) OP for three values of \(h_{\textrm{NTFP}}\). (b) AAR for three values of \(h_{\textrm{NTFP}}\). (c) OP for DT, RT, and HT. (d) AAR for DT, RT, and HT.}
    \label{fig:PerformancevsSDdist} 
\end{figure*}

\begin{figure}
     \centering
  \subfloat[\(R_1\)\label{fig:PoutvsR1_DT_RT_HT}]{%
       \includegraphics[width=0.49\linewidth]{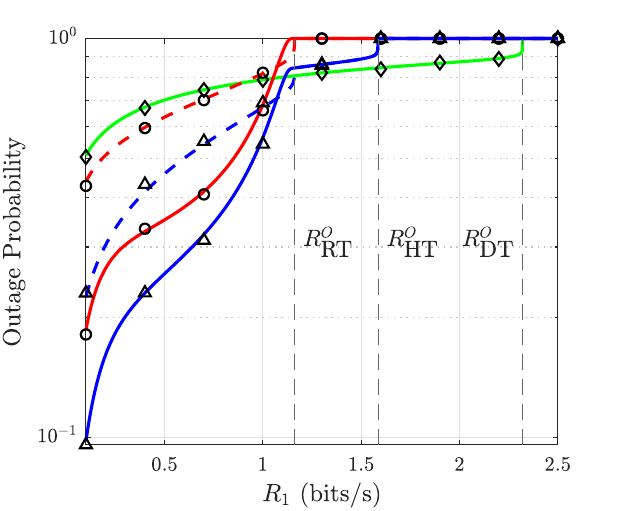}}
  \subfloat[\(R_2\)\label{fig:PoutvsR2_DT_RT_HT}]{%
        \includegraphics[width=0.49\linewidth]{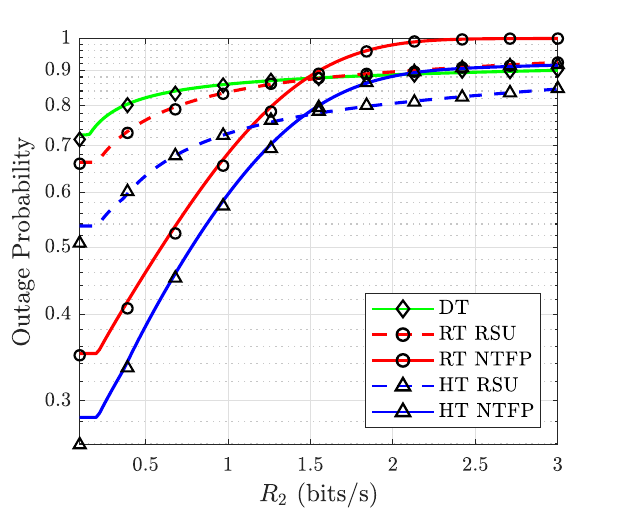}}
  \caption{OP as a function of rates, showing DT, RT, and HT.}
  \label{fig:PerformancevsR1_R2} 
\end{figure}

Fig.~\ref{fig:PerformancevsSDdist} shows the performance as a function of \(||S-D_1||\). Fig.~\ref{fig:PoutvsSD1Distance_NTFPHeight} and Fig.~\ref{fig:AARvsSD1Distance_NTFPHeight} shows the OP and AAR of \(D_1\) for the NTFP and RSU. Note that the relays are always situated in the middle of \(S\) and \(D_1\). Three curves for the NTFP are shown at different heights (\(h_{\textrm{NTFP}}\)). We can see that, depending on \(h_{\textrm{NTFP}}\), the NTFP outperforms the RSU for distances greater than $50$-$100$ m. The RSU achieves better performance for very short distances, but the performance rapidly degrades as the distance increases since LOS becomes increasingly limited. Furthermore, higher values of \(h_{\textrm{NTFP}}\) increases the OP and decreases the AAR. However, the degraded performance is greater for lower \(h_{\textrm{NTFP}}\). This is because, at low heights, building blockages decrease the LOS probability. Fig.~\ref{fig:PoutvsSDdist_DT_RT_HT} and Fig.~\ref{fig:AARvsSDdist_DT_RT_HT} show the OP and AAR of \(D_1\) as a function of \(||S-D_1||\) for DT, RT, and HT. We can see in Fig.~\ref{fig:PoutvsSDdist_DT_RT_HT} that for distances less than $100$ m, using HT with the RSU outperforms the NTFP and gives the best performance. However, for distances between $100$ m and $500$ m, using HT for the NTFP gives the best performance. Since DT is infeasible at distances longer than $500$ m, RT and HT become equivalent. Hence, using the NTFP with RT will give the best performance without adding the complexity of HT. Furthermore, we notice that the difference in performance is minimal between RT and HT for the RSU and that the performance for both cases is improved slightly over DT. Therefore, the relay type and the transmission scheme selection depend on the source and destination distance. From Fig.~\ref{fig:AARvsSDdist_DT_RT_HT}, we see DT outperforming other schemes in terms of AAR for distances less than $400$ m, and the NTFP outperforming for larger distances.

Fig.~\ref{fig:PoutvsR1_DT_RT_HT} compares the performance of the three transmission schemes in terms of OP. We notice that using the NTFP with HT gives the highest performance for \(R_1<1.1~\textrm{bits/s}\), which is due to the fact that both DT and RT links perform well for low rates. At \(R^O_{\textrm{RT}}\), the RT link experiences full outage; therefore, \(S\) can only send directly to \(D_1\), which explains why DT outperforms other schemes at these rates. We note that when RT using NOMA experiences an outage, DT will have a better performance than HT. This is because, unlike HT, DT uses only one time slot. The figure also shows the outage value of HT, \(R^O_{\textrm{HT}}\), and the outage value of DT, \(R^O_{\textrm{DT}}\). We see similar performance for \(D_2\) in Fig.~\ref{fig:PoutvsR2_DT_RT_HT}, however there are no full outages.

\begin{figure}
     \centering
  \subfloat[OP\label{fig:PoutvsNumOfUAV_UAVHeight}]{%
       \includegraphics[width=0.49\linewidth]{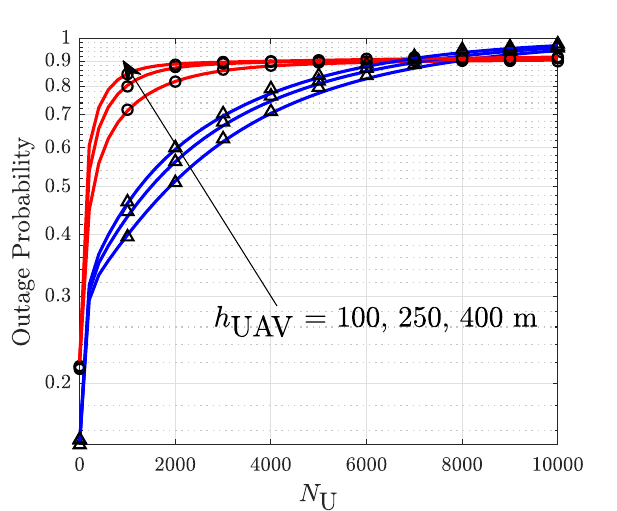}}
  \subfloat[AAR\label{fig:AARvsNumOfUAV_UAVHeight}]{%
        \includegraphics[width=0.49\linewidth]{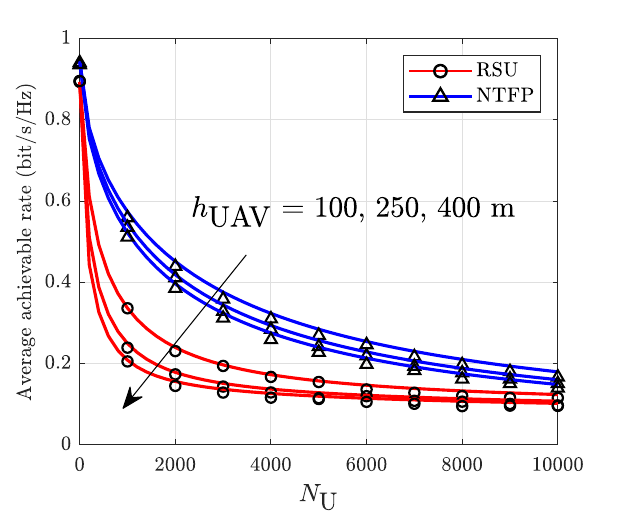}}
  \caption{Performance as a function of \(N_\textrm{U}\) and \(h_{\textrm{UAV}}\).}
  \label{fig:PerformancevsNumOfUAVs} 
\end{figure}

Fig.~\ref{fig:PerformancevsNumOfUAVs} depicts the performance of \(D_1\) as a function of the number of UAVs for three values of \(h_{\textrm{UAV}}\). Fig.~\ref{fig:PoutvsNumOfUAV_UAVHeight} shows the OP while Fig.~\ref{fig:AARvsNumOfUAV_UAVHeight} shows the AAR. We notice that increasing the number of UAVs heavily increases the OP and decreases the AAR of the RSU compared to the NTFP. This is because the NTFP has a higher probability of LOS for the main link, alleviating  the impact of UAV interference. We also notice that increasing the height of the UAVs moderately increases the OP and decreases the AAR for both RSU and NTFP. This can be explained for the RSU because higher altitudes of the UAVs increase the probability of a LOS interference link to the RSU. As for the NTFP, higher altitudes of the UAVs decrease the distance between the UAVs and the NTFP; hence, interference will be more substantial. Finally, we notice how at large \(N_\textrm{U}\) the RSU begins to outperform the NTFP. This is due to the differing probability of LOS models for A2G and A2A links. At a certain point, increasing \(N_\textrm{U}\) does not increase the interference on the RSU significantly. On the other hand, the constant LOS link between the NTFP and the UAVs introduces significant interference, increasing with \(N_\textrm{U}\).

\begin{figure}
     \centering
  \subfloat[NTFP\label{fig:AARvsa1_NTFP}]{%
       \includegraphics[width=0.49\linewidth]{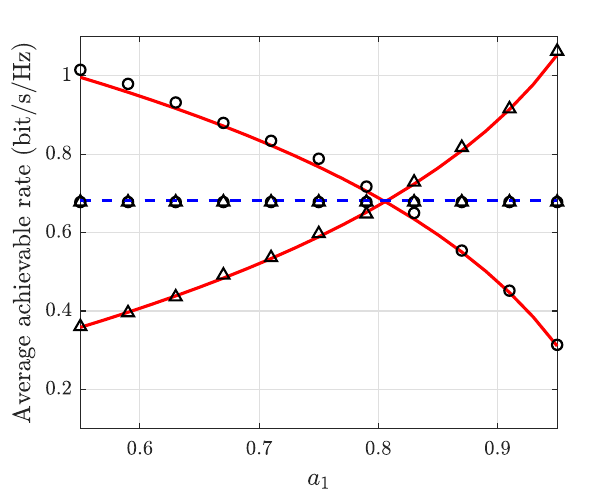}}
  \subfloat[RSU\label{fig:AARvsa1_RSU}]{%
        \includegraphics[width=0.49\linewidth]{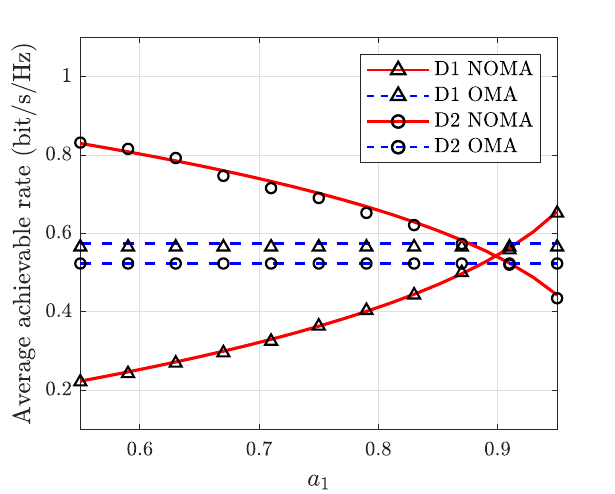}}
  \caption{AAR as a function of \(a_1\), for NTFP and RSU.}
  \label{fig:AARvsa1} 
\end{figure}

\begin{figure}
    \centering
    \subfloat[\(D_1.\)\label{fig:AARD1vsa1_DT_RT_HT}]{%
    \includegraphics[width=0.49\linewidth]{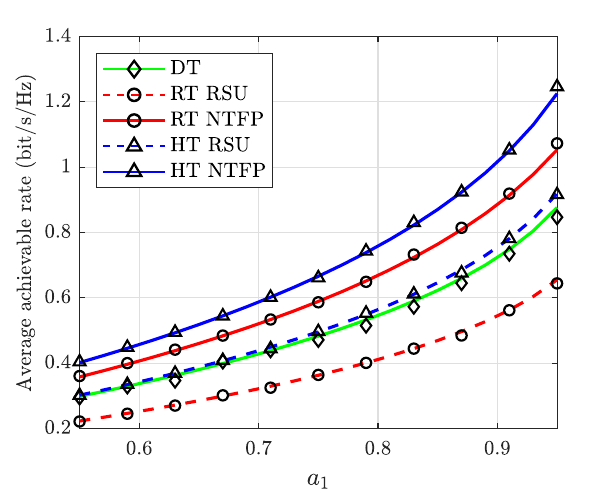}}
    \subfloat[\(D_2.\)\label{fig:AARD2vsa1_DT_RT_HT}]{%
    \includegraphics[width=0.49\linewidth]{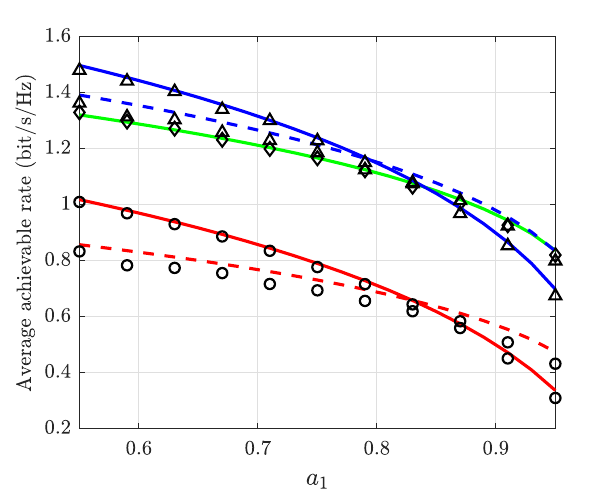}}
    \\
    \subfloat[\(D_1.\)\label{fig:PoutD1vsa1_DT_RT_HT}]{%
    \includegraphics[width=0.49\linewidth]{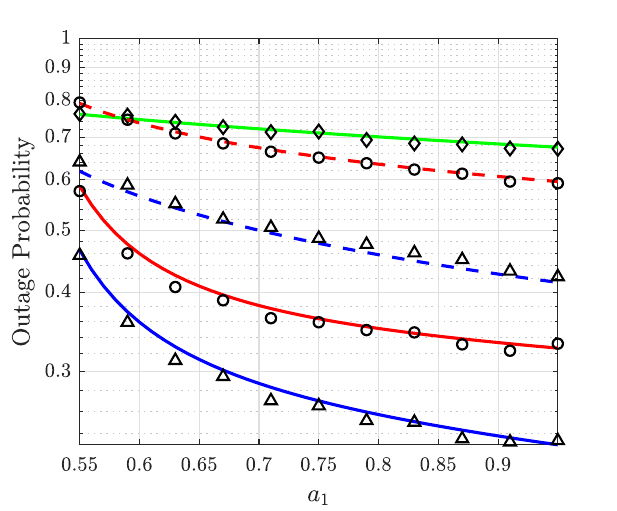}}
    \subfloat[\(D_2.\)\label{fig:PoutD2vsa1_DT_RT_HT}]{%
    \includegraphics[width=0.49\linewidth]{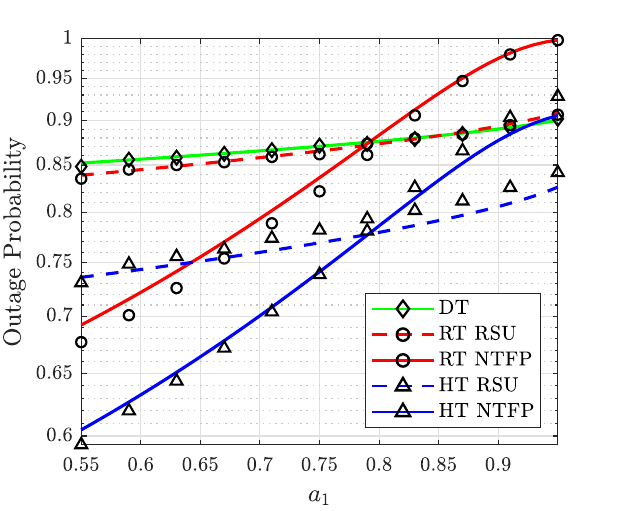}}
    \caption{Performance of \(D_1\) and \(D_2\) as a function of \(a_1\).}
    \label{fig:Performance_vs_a1} 
\end{figure}

\begin{figure*}[ht!]
    \centering
    \subfloat[\(D_1.\)\label{fig:PoutvsR1_DT_RT_HT_2nd_Config}]{%
    \includegraphics[width=0.25\linewidth]{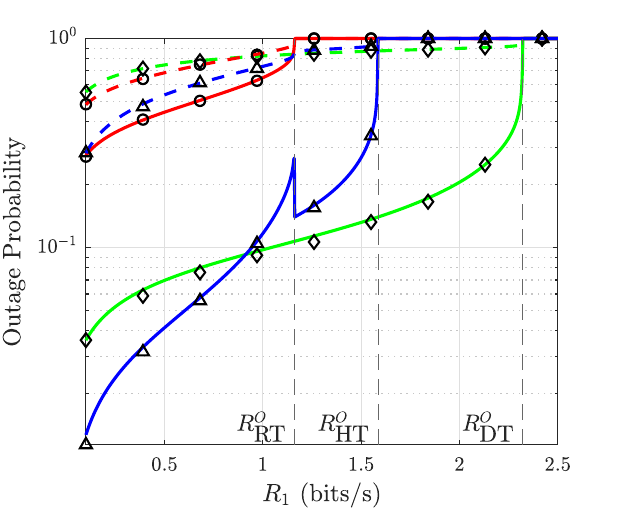}}
    \subfloat[\(D_2.\)\label{fig:PoutvsR2_DT_RT_HT_2nd_Config}]{%
    \includegraphics[width=0.25\linewidth]{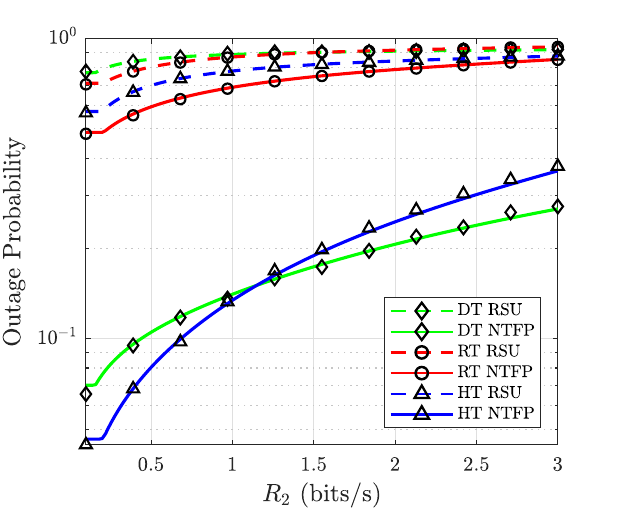}}
    \subfloat[\(D_1.\)\label{fig:PoutvsSD1dist_DT_RT_HT_2nd_Config}]{%
    \includegraphics[width=0.25\linewidth]{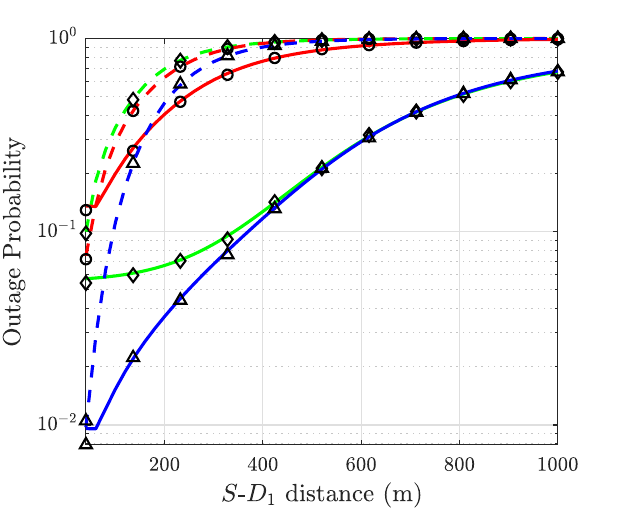}}
    \subfloat[\(D_2.\)\label{fig:PoutvsSD2dist_DT_RT_HT_2nd_Config}]{%
    \includegraphics[width=0.25\linewidth]{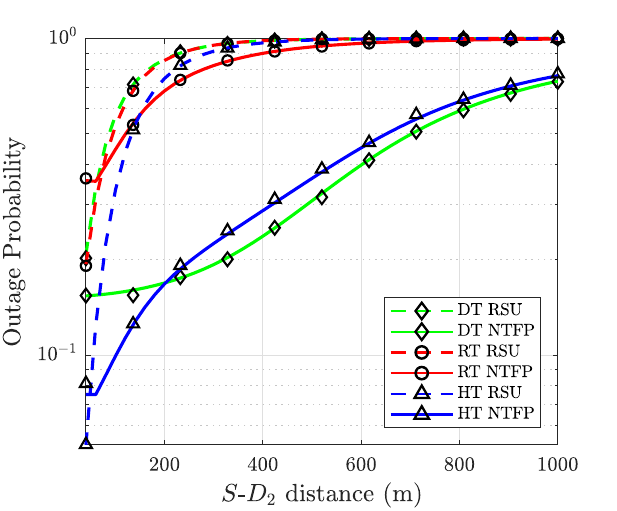}}
    \caption{Performance when NTFP and RSU act as a source.}
    \label{fig:NTFP_RSU_as_Source} 
\end{figure*}

\begin{figure*}[ht!]
    \centering
    \subfloat[\label{fig:PoutvsgammaNOMAvsOMAa1}]{%
    \includegraphics[width=0.25\linewidth]{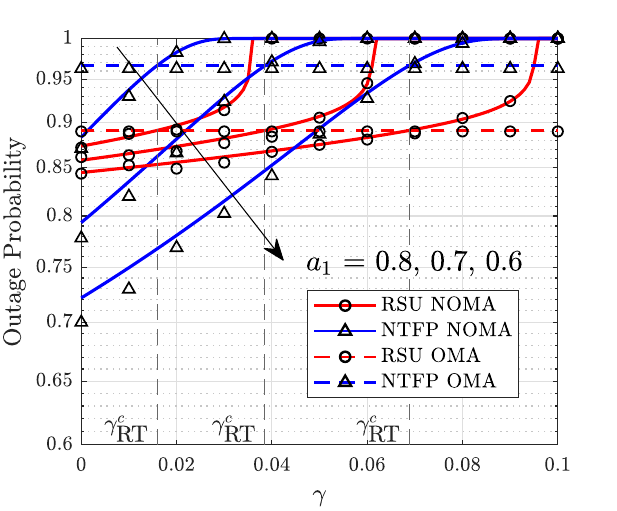}}
    \subfloat[\label{fig:Poutvsgamma_DT_RT_HT}]{%
    \includegraphics[width=0.25\linewidth]{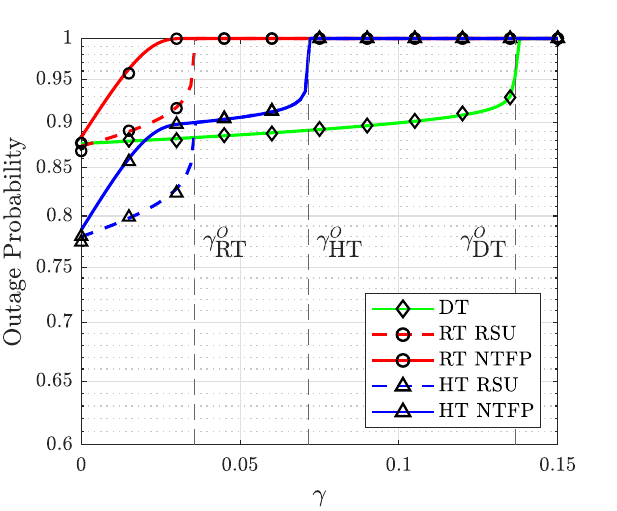}}
    \subfloat[\label{fig:AARvsgamma_a1}]{%
    \includegraphics[width=0.25\linewidth]{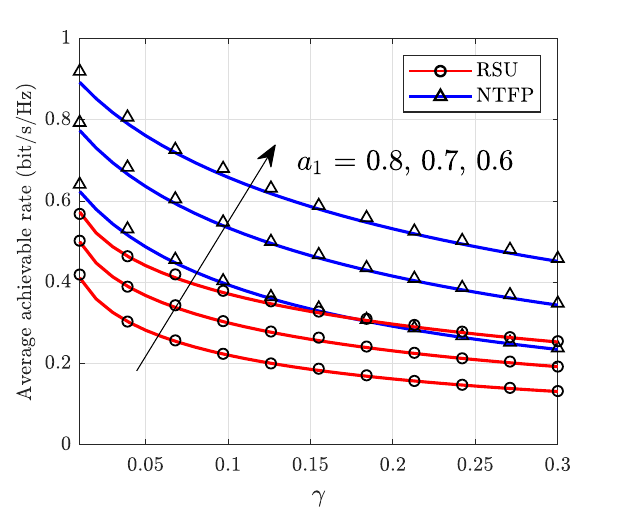}}
    \subfloat[\label{fig:AARvsgamma_DT_RT_HT}]{%
    \includegraphics[width=0.25\linewidth]{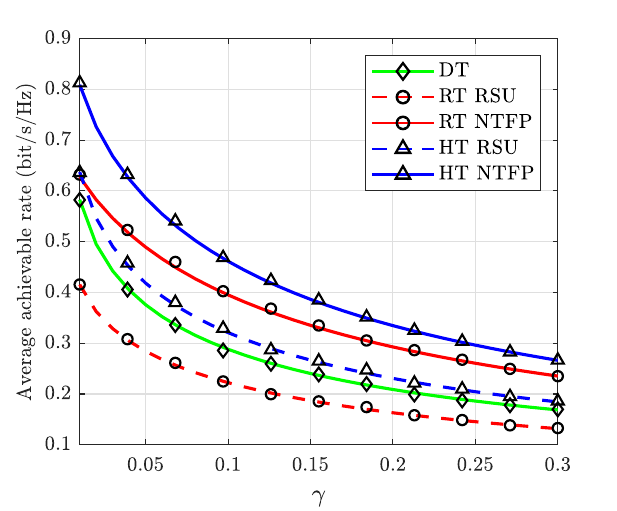}}
    \caption{The effect of imperfect SIC on the performance of \(D_2\). (a) OP vs \(\gamma\) showing three values of \(a_1\). (b) OP vs \(\gamma\) showing DT, RT, and HT. (c) AAR vs \(\gamma\) showing three values of \(a_1\). (d) AAR vs \(\gamma\) showing DT, RT, and HT.}
    \label{fig:SIC_Figures} 
\end{figure*}

\begin{figure}[ht!]
    \centering
  \includegraphics[width=.72\linewidth]{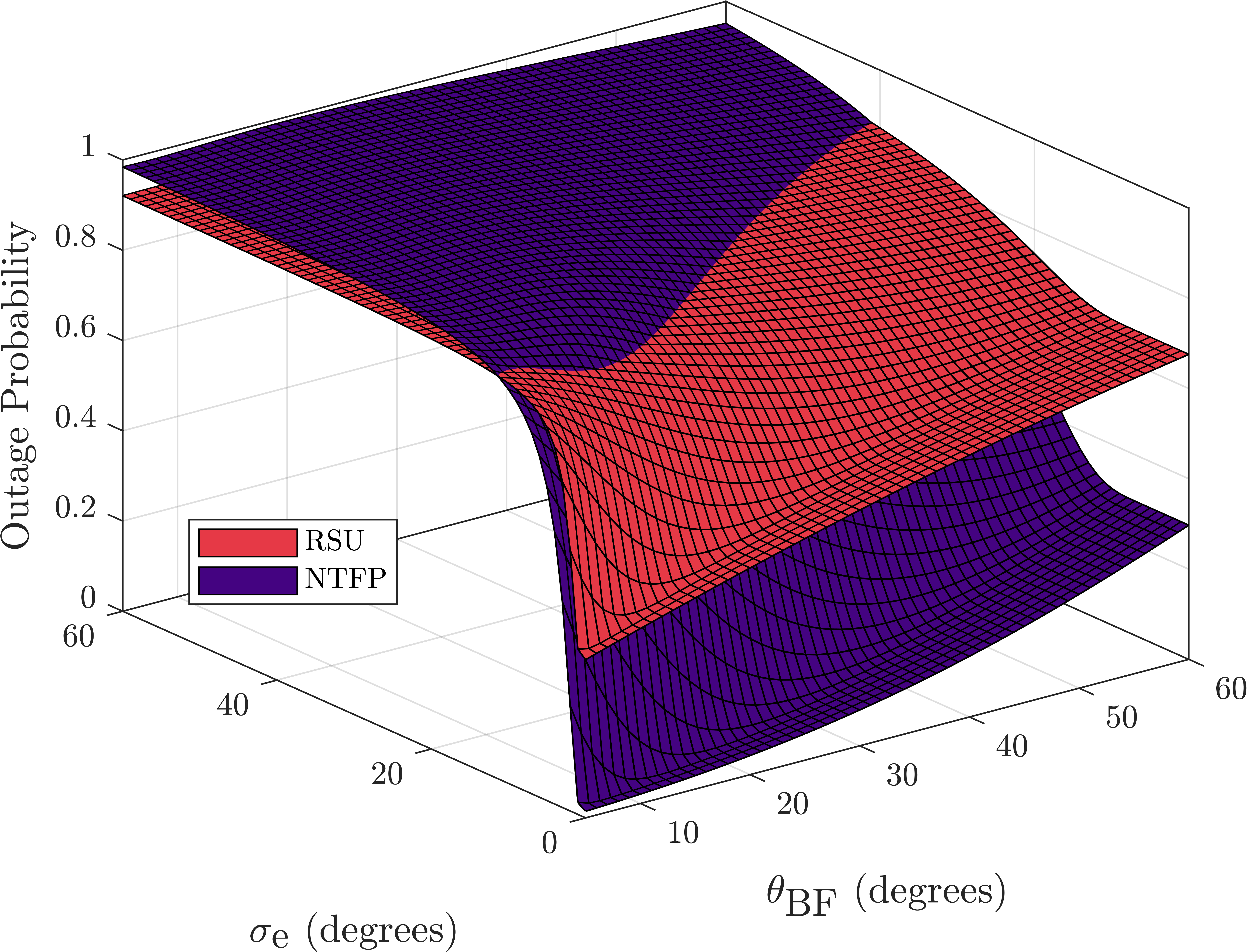}
  \captionof{figure}{OP as a function of \(\sigma_{\textrm{e}}\) and \(\theta_{\textrm{BF}}\).}
  \label{fig:PoutvsSigmaBE_ThetaBF3D}
\end{figure}

\begin{figure}[ht!]
    \centering
  \includegraphics[width=.72\linewidth]{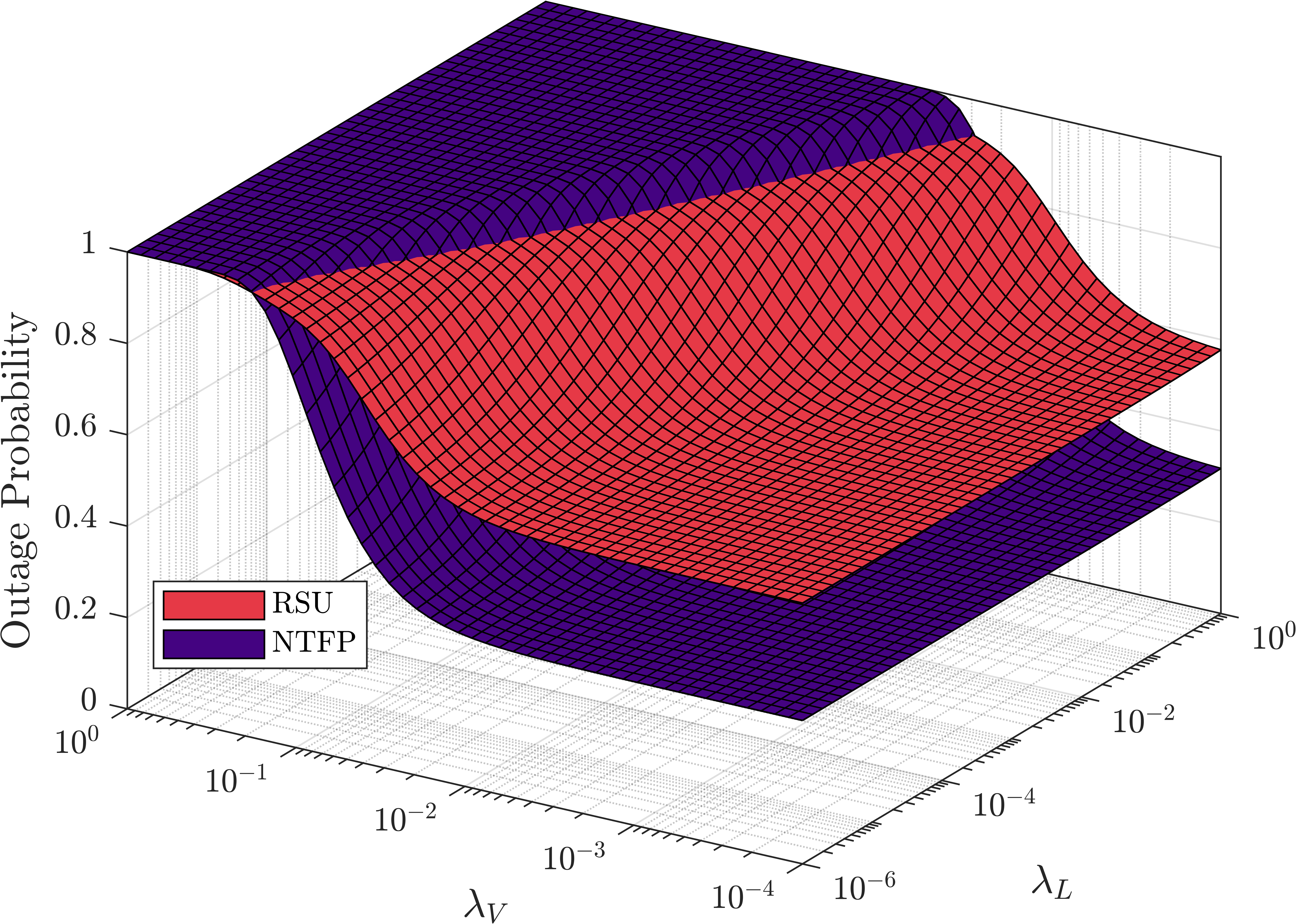}
  \captionof{figure}{OP as a function of \(\lambda_{V}\) and \(\lambda_{L}\).}
  \label{fig:Poutvslambdas3D}
\end{figure}

In Fig.~\ref{fig:AARvsa1}, we show the RT AAR of \(D_1\) and \(D_2\) as a function of \(a_1\), considering OMA and NOMA when using an NTFP and an RSU. Since \(a_1\) is the power fraction of \(D_1\), we see that the AAR of \(D_1\) increases when \(a_1\) increases, while the AAR of \(D_2\) decreases with increasing \(a_1\). We also see that there is a tradeoff between the performance of \(D_1\) and \(D_2\) when using NOMA, where, depending on \(a_1\), only one of the users may outperform its OMA performance. 

Fig.~\ref{fig:Performance_vs_a1} shows the performance in terms of OP and AAR as a function of \(a_1\), for DT, RT, and HT. We see that, for \(D_1\), the NTFP outperforms the RSU and that HT outperforms other schemes. Furthermore, we see that using the RSU does not provide any advantage in terms of AAR compared to DT. For \(D_2\), we notice the NTFP outperforms the RSU for lower values of \(a_1\), and the RSU outperforms for higher values, which is due to the phenomenon explained in Fig~\ref{fig:LOS_NLOS_Total}.

In Fig.~\ref{fig:NTFP_RSU_as_Source}, we study the configuration when NTFP and RSU act as a source when using NOMA. We can see in Fig.~\ref{fig:PoutvsR1_DT_RT_HT_2nd_Config}, which plots the OP as a function of \(R_1\), that the NTFP using HT has the best performance for \(R_1<0.9\) bits/s. This is because the NTFP has a high probability of achieving a LOS link, enabling a strong DT link which is advantageous for HT. However, for rates larger than \(R^O_{\textrm{RT}}\), RT experiences a full outage; hence, DT becomes the best scheme. The pulse in the HT curve occurs due to how HT is defined; when the NTFP-relay link is unavailable (full outage), the NTFP uses DT again in the second time slot; thus, it takes advantage of the strong LOS link to the destination vehicle. In Fig~\ref{fig:PoutvsR2_DT_RT_HT_2nd_Config}, we see similar performance for \(D_2\); however, the links do not experience a full outage. Fig.~\ref{fig:PoutvsSD1dist_DT_RT_HT_2nd_Config} and Fig.~\ref{fig:PoutvsSD2dist_DT_RT_HT_2nd_Config} show that the NTFP outperforms the RSU when the distance increases between the source and destination. Furthermore, there is no advantage to the NTFP using HT over DT. 

Fig.~\ref{fig:SIC_Figures} showcases the effect of imperfect SIC on the performance of the second user, \(D_2\). Fig.~\ref{fig:PoutvsgammaNOMAvsOMAa1} plots the OP to compare between NOMA and OMA. We see that increasing \(\gamma\) increases the NOMA OP for \(D_2\), since the interference increases. We also see that OMA outperforms NOMA for values above the cross-point threshold \(\gamma^c_{\textrm{RT}}\). Fig.~\ref{fig:Poutvsgamma_DT_RT_HT} plots the OP and compares between DT, RT, and HT. We notice how at certain \(\gamma\) values, the different transmission schemes encounter a full outage, and that, as seen before, DT allows for higher values of \(\gamma\) than RT and HT. Fig.~\ref{fig:AARvsgamma_a1} and Fig.~\ref{fig:AARvsgamma_DT_RT_HT} show the effect of imperfect SIC on the AAR. Predictably, the AAR decreases with increasing \(\gamma\). However, we notice that the AAR of the NTFP increases more rapidly than that of the RSU when \(a_1\) is decreased. Finally, in terms of DT, RT, and HT, we notice that the relation between the schemes is unchanged as \(\gamma\) changes.

\begin{table*}[ht!]
\tiny
\caption{Summary of OP analysis results.}
  \label{tab:PoutTable}
  \centering
  \resizebox{\textwidth}{!}
  {%
    \begin{tabular}{ccclclclcl}
    \hline
\multicolumn{1}{l}{}                         & \multicolumn{9}{c}{\(D_1\): 
High Priority and Low Data Rates}    
\\ \hline
\multicolumn{1}{l}{}                         & NOMA Power                            & \multicolumn{4}{c|}{\(a_1<0.7\)}                                                     & \multicolumn{4}{c}{\(a_1>0.7\)}                                                     \\ \hline
\multicolumn{1}{l}{}                         & Distance                              & \multicolumn{2}{c|}{\(||S-D_1||<100\) m} & \multicolumn{2}{c|}{\(||S-D_1||>>100\) m} & \multicolumn{2}{c|}{\(||S-D_1||<100\) m} & \multicolumn{2}{c}{\(||S-D_1||>>100\) m} \\ \hline
\multicolumn{1}{c|}{\multirow{2}{*}{Relay}}  & \multicolumn{1}{c|}{\(R_1<1\) bits/s} & \multicolumn{2}{c|}{RSU, HT}             & \multicolumn{2}{c|}{NTFP, RT}             & \multicolumn{2}{c|}{RSU, HT}             & \multicolumn{2}{c}{NTFP, RT}             \\ \cline{2-2}
\multicolumn{1}{c|}{}                        & \multicolumn{1}{c|}{\(R_1>1\) bits/s} & \multicolumn{2}{c|}{DT}                  & \multicolumn{2}{c|}{DT}                   & \multicolumn{2}{c|}{DT}                  & \multicolumn{2}{c}{DT}                   \\ \hline
\multicolumn{1}{c|}{\multirow{2}{*}{Source}} & \multicolumn{1}{c|}{\(R_1<1\) bits/s} & \multicolumn{2}{c|}{NTFP, HT}            & \multicolumn{2}{c|}{NTFP, DT}             & \multicolumn{2}{c|}{NTFP, HT}            & \multicolumn{2}{c}{NTFP, DT}             \\ \cline{2-2}
\multicolumn{1}{c|}{}                        & \multicolumn{1}{c|}{\(R_1>1\) bits/s} & \multicolumn{2}{c|}{NTFP, DT}            & \multicolumn{2}{c|}{NTFP, DT}             & \multicolumn{2}{c|}{NTFP, DT}            & \multicolumn{2}{c}{NTFP, DT}             \\ \hline
                                             & \multicolumn{9}{c}{\(D_2\): Low Priority and High Data Rates}                                                                                                                                                                            \\ \hline
                                             & NOMA Power                            & \multicolumn{4}{c|}{\(a_2<0.3\)}                                                     & \multicolumn{4}{c}{\(a_2>0.3\)}                                                     \\ \hline
                                             & Distance                              & \multicolumn{2}{c|}{\(||S-D_2||<100\) m} & \multicolumn{2}{c|}{\(||S-D_2||>>100\) m} & \multicolumn{2}{c|}{\(||S-D_2||<100\) m} & \multicolumn{2}{c}{\(||S-D_2||>>100\) m} \\ \hline
\multicolumn{1}{c|}{\multirow{2}{*}{Relay}}  & \multicolumn{1}{c|}{\(R_2<2\) bits/s} & \multicolumn{2}{c|}{RSU, HT}             & \multicolumn{2}{c|}{NTFP, RT}             & \multicolumn{2}{c|}{RSU, HT}             & \multicolumn{2}{c}{NTFP, RT}             \\ \cline{2-2}
\multicolumn{1}{c|}{}                        & \multicolumn{1}{c|}{\(R_2>2\) bits/s} & \multicolumn{2}{c|}{RSU, HT}             & \multicolumn{2}{c|}{RSU, HT}              & \multicolumn{2}{c|}{RSU, HT}             & \multicolumn{2}{c}{NTFP, HT}             \\ \hline
\multicolumn{1}{c|}{\multirow{2}{*}{Source}} & \multicolumn{1}{c|}{\(R_2<2\) bits/s} & \multicolumn{2}{c|}{NTFP, HT}            & \multicolumn{2}{c|}{NTFP, DT}             & \multicolumn{2}{c|}{NTFP, HT}            & \multicolumn{2}{c}{NTFP, DT}             \\ \cline{2-2}
\multicolumn{1}{c|}{}                        & \multicolumn{1}{c|}{\(R_2>2\) bits/s} & \multicolumn{2}{c|}{NTFP, HT}            & \multicolumn{2}{c|}{NTFP, DT}             & \multicolumn{2}{c|}{NTFP, HT}            & \multicolumn{2}{c}{NTFP, DT}             \\ \hline
\end{tabular}}%
\end{table*}

\begin{table*}[h!]
\tiny
  \centering
  \caption{Summary of AAR analysis results.}
  \label{tab:AARTable}
  \resizebox{\textwidth}{!}{%
    \begin{tabular}{cclclclcl}
\hline
\multicolumn{9}{c}{\(D_1\): High Priority and Low Data Rates}                                                                                                                                                              \\ \hline
NOMA Power                  & \multicolumn{4}{c|}{\(a_1<0.7\)}                                                     & \multicolumn{4}{c}{\(a_1>0.7\)}                                                     \\ \hline
Distance                    & \multicolumn{2}{c|}{\(||S-D_1||<100\) m} & \multicolumn{2}{c|}{\(||S-D_1||>>100\) m} & \multicolumn{2}{c|}{\(||S-D_1||<100\) m} & \multicolumn{2}{c}{\(||S-D_1||>>100\) m} \\ \hline
\multicolumn{1}{c|}{Relay}  & \multicolumn{2}{c|}{DT}                  & \multicolumn{2}{c|}{NTFP, RT}             & \multicolumn{2}{c|}{DT}                  & \multicolumn{2}{c}{NTFP, RT}             \\ \hline
\multicolumn{1}{c|}{Source} & \multicolumn{2}{c|}{NTFP, DT}            & \multicolumn{2}{c|}{NTFP, DT}             & \multicolumn{2}{c|}{NTFP, DT}            & \multicolumn{2}{c}{NTFP, DT}             \\ \hline
\multicolumn{9}{c}{\(D_2\): Low Priority and High Data Rates}                                                                                                                                                                  \\ \hline
NOMA Power                  & \multicolumn{4}{c|}{\(a_2<0.3\)}                                                     & \multicolumn{4}{c}{\(a_2>0.3\)}                                                     \\ \hline
Distance                    & \multicolumn{2}{c|}{\(||S-D_2||<100\) m} & \multicolumn{2}{c|}{\(||S-D_2||>>100\) m} & \multicolumn{2}{c|}{\(||S-D_2||<100\) m} & \multicolumn{2}{c}{\(||S-D_2||>>100\) m} \\ \hline
\multicolumn{1}{c|}{Relay}  & \multicolumn{2}{c|}{DT}                  & \multicolumn{2}{c|}{NTFP, RT}             & \multicolumn{2}{c|}{DT}                  & \multicolumn{2}{c}{NTFP, RT}             \\ \hline
\multicolumn{1}{c|}{Source} & \multicolumn{2}{c|}{RSU, DT}             & \multicolumn{2}{c|}{NTFP, DT}             & \multicolumn{2}{c|}{RSU, DT}             & \multicolumn{2}{c}{NTFP, DT}             \\ \hline
\end{tabular}}%
\end{table*}

Fig.~\ref{fig:PoutvsSigmaBE_ThetaBF3D} shows the OP of \(D_1\) as a function of the beamforming angle and the beam steering error. We see that increasing \(\sigma_e\) increases the OP. However, for small values of \(\sigma_e\), the outage remains constant since the error is not large enough to exceed \(\theta_{\textrm{BF}}\). Therefore, smaller beam width angles are more prone to steering errors. Furthermore, we notice that increasing \(\theta_{\textrm{BF}}\) increases the OP. This is because the destination node will experience stronger interference from other nodes for higher beam width angles. The figure shows that for a given \(\sigma_e\), there exists a value of \(\theta_{\textrm{BF}}\) that minimizes the OP. Furthermore, the NTFP has a high probability of LOS; therefore, higher values of \(\sigma_e\) and \(\theta_{\textrm{BF}}\) will negatively impact its performance since the interference will also be in LOS. Consequently, the RSU outperforms the NTFP for higher values of \(\sigma_e\) and \(\theta_{\textrm{BF}}\).

Fig.~\ref{fig:Poutvslambdas3D} draws the OP of \(D_1\) as a function of \(\lambda_{\textrm{V}}\) and \(\lambda_{\textrm{L}}\). We see that, for low intensity of vehicles and roads, the performance is constant and affected only by the interfering UAVs. High values of \(\lambda_{\textrm{V}}\) correspond to congested roads with significant interference; hence, the OP increases drastically for the NTFP and RSU. As seen in previous figures, when the interference sources increase drastically, the RSU begins to outperform the NTFP.

\section{Summary and Conclusion} \label{conclusion}
A summary is given in this section to provide a complete perspective on the system's performance. We show in Table~\ref{tab:PoutTable} and Table~\ref{tab:AARTable} the combination of infrastructure and transmission scheme choices with the highest performance in terms of OP and AAR for a given rate, NOMA power, and communication distance. The tables summarize the results for \(D_1\) and \(D_2\) and the two studied configurations. The following recommendations are made for a system design. First, when NTFP and RSU act as a source, it is best to use the NTFP. However, when NTFP and RSU act as a relay, short distances are better served by the RSU, whereas the NTFP better serves long distances. Furthermore, when the objective is to reach the highest attainable data rates possible, DT should be used since it has the largest NOMA outage rate, i.e. \(R^O_{\textrm{DT}}>R^O_{\textrm{HT}}>R^O_{\textrm{RT}}\). Moreover, the power coefficient \(a_1\) determines the performance cross-point between NOMA and OMA. Therefore, it should be chosen according to system requirements and specifications. Larger values of \(a_1\) allow for a larger range of data rates for \(D_1\) such that NOMA outperforms OMA. Conversely, for \(D_2\), larger values of \(a_1\) decrease the range of rates such that NOMA outperforms OMA. With regards to imperfect SIC, the results predictably showed decreased performance for higher SIC errors. It was also shown that DT has the highest tolerance of SIC errors, i.e. \(\gamma^O_{\textrm{DT}}>\gamma^O_{\textrm{HT}}>\gamma^O_{\textrm{RT}}\). Finally, \(\theta_{\textrm{BF}}\) must be chosen carefully according to which infrastructure is used and the value of \(\sigma_e\). Specifically, the NTFP relies on a LOS link for better performance, and thus, underperforms compared to RSU for higher errors.

In summary, we comparatively studied the performance of NTFPs and RSUs in mmWave VANETs in dense urban environments. We derived approximate expressions for the OP and AAR in the presence of interference from UAM aircraft using DT, RT, and HT schemes. Depending on the type of link, two LOS probability models were implemented. Furthermore, NOMA was employed and compared with OMA. Results showed that NTFPs outperform RSUs when vehicles are spatially distant and that the optimal transmission scheme largely depended on the distance and NOMA power coefficients. The insights from our paper can be used in VANETs to design mechanisms that opportunistically choose the most suitable platform and type of communication in concordance with system setup planning requirements.


%

\appendices
\vspace{-0.3cm}
\section{Proof of Lemma 1} \label{AppendixLemma1}
To calculate the Laplace transform of \(I_{L_0}\), we start with the definition of a Laplace transform
\begin{equation}
    \begin{split}
        \mathcal{L}_{I_{L_0}}(s) = \mathbb{E}[\textrm{e}^{-sI_{L_0}}]
        & 
        = \mathbb{E}\left[\prod_{v\in\Phi_{\textrm{V}}}\textrm{exp}\left(-s|h_{v q}|^2l_{v q}\right)\right]\\
        &\!\!\!\!\!\!\!\!\!\!\!\!\!\!\!\!\!\!\!\!\!\!\!\!\!\!\!\!\!\!\!\!\!\!\!\! \stackrel{(a)}{=} \mathbb{E}\left[\prod_{v\in\Phi_{\textrm{V}}}\mathbb{E}_{|h_{v q}|^2}\Big[\textrm{exp}\left(-s|h_{v q}|^2l_{v q}\right)\Big]\right]\\
        &\!\!\!\!\!\!\!\!\!\!\!\!\!\!\!\!\!\!\!\!\!\!\!\!\!\!\!\!\!\!\!\!\!\!\!\! \stackrel{(b)}{=} \mathbb{E}\left[\prod_{v\in\Phi_{\textrm{V}}}\left(\frac{m}{s l_{v q}+m}\right)^m\right],
    \end{split}
\end{equation}
where \((a)\) follows from the fact that the coefficients \(|h_{v q}|^2\) are independent, and \((b)\) is an established result for Nakagami-$m$. To obtain Eq.~\eqref{LT_IL0}, we use the probability generating functional (PGFL) of a PPP \cite{andrews2016primer}.

\vspace{-0.3cm}
\section{Proof of Lemma 2} \label{AppendixLemma2}
We calculate the Laplace transform of \(I_{L}\) as follows
\begin{equation}
    \begin{split}
        \mathcal{L}_{I_{L}}(s) & = \mathbb{E}\left[\textrm{exp}\left(-s\sum_{L_j\in\Phi_{L}}\sum_{v_{L_j}\in\Phi_{\textrm{V}_{q}}} |h_{v_{L_j} q}|^2l_{v_{L_j} q}\right)\right]\\
        & = \mathbb{E}\left[\prod_{L_j\in\Phi_{L}}\prod_{v_{L_j}\in\Phi_{\textrm{V}_{q}}}\textrm{exp}\left(-s|h_{v_{L_j} q}|^2l_{v_{L_j} q}\right)\right].
    \end{split}
\end{equation}
\indent Following the steps in Appendix~\ref{AppendixLemma1}, and using the PGFL of a 1D PPP, we get
\begin{equation}
    \begin{split}
        \mathcal{L}_{I_{L}}(s) & =\mathbb{E}\Bigg[\prod_{L_j\in\Phi_{L}}\textrm{exp}\Bigg(-\lambda_{\textrm{V}} \int_{\mathbb{R}}1\\
        & -\left(\frac{m}{s (x^2+{y_j}^2+z^2)^{-\frac{\alpha}{2}}+m}\right)^m\textrm{d}x\Bigg)\Bigg].
    \end{split}
\end{equation}
\indent Now  we use the PGFL of the 2D PPP, \(\Phi_L\), which has the representation space \(\mathbb{R}^+ \times [0,2\pi)\) \cite{Cox2020Book} and we obtain Eq.~\eqref{LT_IL}.
\vspace{-0.3cm}
\section{Proof of Lemma 3} \label{AppendixLemma3}
To calculate the Laplace transform of the 2D BPP, we proceed as follows
\begin{equation} \label{LTUAVBPP}
    \begin{split}
        \mathcal{L}_{I_{\textrm{U}}}(s) & 
        =\mathbb{E}\left[\prod_{N_{\textrm{U}}}\textrm{exp}\left(-s|h_{u q}|^2l_{u q}\right)\right]\\
        &\!\!\!\!\!\!\!\!\!\!\!\!\!\!\!=\mathbb{E}\left[\prod_{N_{\textrm{U}}}\mathbb{E}_{|h_{u q}|^2}\left[\textrm{exp}\left(-s|h_{u q}|^2l_{u q}\right)\right]\right]\\
        &\!\!\!\!\!\!\!\!\!\!\!\!\!\!\!=\mathbb{E}_{l_{u q}}\left[\prod_{N_{\textrm{U}}}\left(\frac{m}{s l_{u q}+m}\right)^m\right]={\mathbb{E}_{l_{u q}}\left[\left(\frac{m}{s l_{u q}+m}\right)^m\right]}^{N_{\textrm{U}}}.
    \end{split}
\end{equation}
\indent To compute the expectation in Eq.~\eqref{LTUAVBPP}, we use \cite{chetlur2017BPP} 
\begin{equation}
    f_{\textrm{U}_x}(u_x)=
    \small
    \begin{cases}
        \frac{2u_x}{L^2}, & \Delta h<u_x<\sqrt{{\Delta h}^2+L^2},\\
        0, & \textrm{otherwise}.
    \end{cases}
    \normalsize
\end{equation}
\indent Hence, we get the Laplace transform in Eq.~\eqref{UAVLaplace}.

\vspace{-0.5cm}
\section{Proof of Theorem 1} \label{AppendixTheorem1}
\indent To calculate the OP of DT for \(D_1\), we proceed as follows
\begin{equation}
    \begin{split}
        P(O_{D_1}^{(\mathbf{DT})}) & = 1-P({O_{D_1}^{(\mathbf{DT})}}^C)\\
    &\!\!\!\!\!\!\!\!\!\!\!\!\!\!\!\!\!\!\!\!\!\!\!\!\!\!\! = 1-\mathbb{E}_{I_{\textrm{tot}_{D_1}}}\left[P\left(|h_{SD_1}|^2>\frac{T_1I_{\textrm{tot}_{D_1}}}{\left(a_1-T_1 a_2\right)l_{SD_1}}\right)\right].
    \end{split}
\end{equation}

Since \(|h_{SD_1}|^2\) is a gamma random variable, we can write the probability as

\begin{equation} \label{DT_NOMA_D1}
   P({O_{D_1}^{(\mathbf{DT})}}^C) = \mathbb{E}_{I_{\textrm{tot}_{D_1}}}\left[\frac{\Gamma\left(m,\frac{mG_1}{l_{SD_1}}I_{\textrm{tot}_{D_1}}\right)}{\Gamma\left(m\right)}\right],
\end{equation}
where \(G_1=T_1/(a_1-T_1 a_2)\). Then, for a constant \(z>0\), we use the following bound \cite{alzer1997some}
\begin{equation}
    \frac{\Gamma(m,z)}{\Gamma(z)} > (1-\textrm{exp}(-cz))^m,
\end{equation}
where \(\Gamma(.,.)\) is the upper incomplete gamma function and \(c=\Gamma(m+1)^{-\frac{1}{m}}\) for \(m>1\). For low values of \(m\), the lower bound offers a close approximation. Hence, we get
\begin{equation} \label{DT_NOMA_Alzers}
    \begin{split}
        P({O_{D_1}^{(\mathbf{DT})}}^C) & = \mathbb{E}_{I_{\textrm{tot}_{D_1}}}\left[\frac{\Gamma\left(m,sI_{\textrm{tot}_{D_1}}\right)}{\Gamma\left(m\right)}\right]\\
        &\!\!\!\!\!\!\!\!\!\!\!\!\!\!\!\!\!\! \gtrapprox \mathbb{E}_{I_{\textrm{tot}_{D_1}}}\left[(1-\textrm{exp}(-csI_{\textrm{tot}_{D_1}}))^m\right]\\
        &\!\!\!\!\!\!\!\!\!\!\!\!\!\!\!\!\!\! = \mathbb{E}_{I_{\textrm{tot}_{D_1}}}\left[\sum_{k=0}^m (-1)^k \binom{m}{k} \textrm{exp}(-cksI_{\textrm{tot}_{D_1}})\right]\\
        &\!\!\!\!\!\!\!\!\!\!\!\!\!\!\!\!\!\! = \sum_{k=0}^m (-1)^k \binom{m}{k} \mathbb{E}\left[\textrm{exp}(-cks(I_{V}+I_{U}))\right]\\
        &\!\!\!\!\!\!\!\!\!\!\!\!\!\!\!\!\!\! = \sum_{k=0}^m (-1)^k \binom{m}{k} \mathcal{L}_{I_{V}}(cks) \mathcal{L}_{I_{U}}(cks)\\
        &\!\!\!\!\!\!\!\!\!\!\!\!\!\!\!\!\!\! = \sum_{k=0}^m (-1)^k \binom{m}{k} \mathcal{H}_{D_1}(cks),
    \end{split}
\end{equation}
\noindent where \(s=\frac{mG_1}{l_{SD_1}}\). Next, we compute the OP for \(D_2\), which is calculated as
\begin{equation}
    \begin{split}
        P(O_{D_2}^{(\mathbf{DT})}) & = 1-P(\textrm{SIR}_{SD_{2\rightarrow 1}}>T_1 \cap \textrm{SIR}_{SD_{2\rightarrow 2}}>T_2)\\
        &\!\!\!\!\!\!\!\!\!\!\!\!\!\!\!\!\!\! = 1-P\Big(|h_{SD_2}|^2>\frac{G_1I_{\textrm{tot}_{D_2}}}{l_{SD_2}}\cap |h_{SD_2}|^2>\frac{G_2I_{\textrm{tot}_{D_2}}}{l_{SD_2}}\Big)\\
        &\!\!\!\!\!\!\!\!\!\!\!\!\!\!\!\!\!\! = 1-P\Big(|h_{SD_2}|^2>\frac{G_{\textrm{max}}I_{\textrm{tot}_{D_2}}}{l_{SD_2}}\Big)\\
        &\!\!\!\!\!\!\!\!\!\!\!\!\!\!\!\!\!\! \cong 1 - \sum_{k=0}^m (-1)^k \binom{m}{k} \mathcal{H}_{D_2}(ck\frac{mG_{\textrm{max}}}{l_{SD_2}}).
    \end{split}
\end{equation}

\section{Proof of Theorem 2} \label{AppendixTheorem2}
\indent To calculate the OP of RT for \(D_1\), we proceed as follows
\begin{equation}
    \begin{split}
        P(O_{D_1}^{(\mathbf{RT})}) & = 1-P({O_{D_1}^{(\mathbf{RT})}}^C) = 1-P(\textrm{SIR}_{SR_1}>T_1^{(\mathbf{RT})})\\
        & \times P(\textrm{SIR}_{RD_{1\rightarrow1}}>T_1^{(\mathbf{RT})}).
    \end{split}
\end{equation}
\indent Then, we follow the same steps as in Appendix \ref{AppendixTheorem1}, then, we obtain Eq.~\eqref{PoutD1_RT_NOMA}. For \(D_2\), we proceed as follows
\begin{equation}
    \begin{split}
        P(O_{D_2}^{(\mathbf{RT})}) & = 1-P({O_{D_2}^{(\mathbf{RT})}}^C)\\
        &\!\!\!\!\!\!\!\!\!\!\!\!\!\!\! = 1-P\Big(\textrm{SIR}_{SR_1}>T_1^{(\mathbf{RT})} \cap \textrm{SIR}_{SR_2}>T_2^{(\mathbf{RT})}\Big) \\
        &\!\!\!\!\!\!\!\!\!\!\!\!\!\!\! \times P\Big(\textrm{SIR}_{RD_{2\rightarrow1}}>T_1^{(\mathbf{RT})} \cap \textrm{SIR}_{RD_{2\rightarrow2}}>T_2^{(\mathbf{RT})}\Big)\\
        &\!\!\!\!\!\!\!\!\!\!\!\!\!\!\! = 1-P\Big(|h_{SR}|^2>\frac{G_{\textrm{max}}^{(\mathbf{RT})}I_{\textrm{tot}_{R}}}{l_{SR}}\Big)\\
        &\!\!\!\!\!\!\!\!\!\!\!\!\!\!\! \times P\Big(|h_{RD_2}|^2>\frac{G_{\textrm{max}}^{(\mathbf{RT})}I_{\textrm{tot}_{D_2}}}{l_{RD_2}}\Big).
    \end{split}
\end{equation}
\indent Following the same steps as in Appendix~\ref{AppendixTheorem1}, we get Eq.~\eqref{PoutD2_RT_NOMA}.
\vspace{-0.3cm}
\section{Proof of Theorem 3} \label{AppendixTheorem3}
To calculate the OP of HT for \(D_1\), we proceed as follows
\begin{equation}
    \begin{split}
        P(O_{D_1}^{(\mathbf{HT})}) & \triangleq P(\textrm{SIR}_{SD_{1\rightarrow1}}<\frac{T_1^{(\mathbf{RT})}}{2} \cap  \textrm{SIR}_{SR_1}<T_1^{(\mathbf{RT})})\\
        &\!\!\!\!\!\!\!\!\!\!\!\!\!\!\!\!\!\! + P(\textrm{SIR}_{SR_1}>T_1^{(\mathbf{RT})}\cap  \textrm{SIR}_{RD_{1\rightarrow1}}<T_1^{(\mathbf{RT})}\\
        &\!\!\!\!\!\!\!\!\!\!\!\!\!\!\!\!\!\! \cap \textrm{SIR}_{SD_{1\rightarrow1}}<T_1^{(\mathbf{RT})}) = 1-P(\textrm{SIR}_{SD_{1\rightarrow1}}>\frac{T_1^{(\mathbf{RT})}}{2})\\
        &\!\!\!\!\!\!\!\!\!\!\!\!\!\!\!\!\!\! -P(\textrm{SIR}_{SR_1}>T_1^{(\mathbf{RT})}) + P(\textrm{SIR}_{SD_{1\rightarrow1}}>\frac{T_1^{(\mathbf{RT})}}{2}\\
        &\!\!\!\!\!\!\!\!\!\!\!\!\!\!\!\!\!\! \cap \textrm{SIR}_{SR_1}>T_1^{(\mathbf{RT})}) + P(\textrm{SIR}_{SR_1}>T_1^{(\mathbf{RT})}\\
        &\!\!\!\!\!\!\!\!\!\!\!\!\!\!\!\!\!\! \cap \textrm{SIR}_{RD_{1\rightarrow1}}<T_1^{(\mathbf{RT})} \cap \textrm{SIR}_{SD_{1\rightarrow1}}<T_1^{(\mathbf{RT})}).
    \end{split}
\end{equation}
\indent The terms \(P(\textrm{SIR}_{SD_{1\rightarrow1}}>\frac{T_1^{(\mathbf{RT})}}{2})\) and \(P(\textrm{SIR}_{SR_1}>T_1^{(\mathbf{RT})})\) are calculated as in Appendix~\ref{AppendixTheorem1}, and the term \(P(\textrm{SIR}_{SD_{1\rightarrow1}}>\frac{T_1^{(\mathbf{RT})}}{2} \cap \textrm{SIR}_{SR_1}>T_1^{(\mathbf{RT})})\) is calculated as in Appendix~\ref{AppendixTheorem2}. To obtain the last term, denoted by \(\mathcal{P}_0\), we proceed as follows
\begin{equation}
    \begin{split}
        \mathcal{P}_0 & = P(\textrm{SIR}_{SR_1}>T_1^{(\mathbf{RT})}) \left(1-P(\textrm{SIR}_{RD_{1\rightarrow1}}>T_1^{(\mathbf{RT})})\right)\\
        & \times \left(1-P(\textrm{SIR}_{SD_{1\rightarrow1}}>T_1^{(\mathbf{RT})})\right),
    \end{split}
\end{equation}
where the three probabilities are calculated in the same was as in Appendix~\ref{AppendixTheorem1}. The same manipulations are made to get the expression for \(D_2\), therefore, they are omitted.






\ifCLASSOPTIONcaptionsoff
  \newpage
\fi



%

\bibliographystyle{IEEEtran}
\bibliography{refs_Abbrev.bib}

\end{document}